\documentclass[lettersize,journal]{IEEEtran}
\usepackage{amsmath,amsfonts}
\usepackage{array}
\usepackage[caption=false,font=normalsize,subrefformat=parens]{subfig}
\usepackage{textcomp}
\usepackage{stfloats}
\usepackage{url}
\usepackage{verbatim}
\usepackage{graphicx}
\usepackage{cite}
\usepackage{algorithm}
\usepackage{algpseudocode}
\usepackage{xcolor}

\begin{document}

\title{Integrated Coexistence for Satellite and Terrestrial Networks \\with Multistatic ISAC}

\author{Jeongju Jee and Jeffrey G. Andrews

\thanks{The authors are with 6G@UT Research Center in the Wireless
Networking and Communications Group, University of Texas
at Austin, Austin, TX 78712, USA
(e-mail: jeongju.jee@utexas.edu; jandrews@ece.utexas.edu).}
}

\maketitle

\begin{abstract}
Tightly integrated low earth orbit (LEO) satellite communications and terrestrial integrated sensing and communication (ISAC) are expected to be key novel aspects of the 6G era. Spectrum sharing between satellite and terrestrial cellular networks may, however, cause severe interference. This paper introduces a cooperation framework for integrated coexistence between satellite and terrestrial networks where the terrestrial network also deploys multistatic ISAC.  Unlike prior works that assume ideal channel state information (CSI) acquisition, the proposed approach develops a practical structure consisting of pre-optimization and refinement stages that leverages the predictability of satellite CSI. In addition, a co-design of terrestrial beamforming and satellite power allocation utilizing a weighted minimum mean-squared error algorithm is proposed, and a target-radar association method designed for multistatic ISAC is presented. Simulation results show that the proposed approach significantly enhances the performance of these integrated networks. Furthermore, it is confirmed that the overall performance approaches the interference-free benchmark as the number of spot beams and radar receivers increases, demonstrating the feasibility of spectral coexistence between the two networks.
\end{abstract}


\section{Introduction}
\IEEEPARstart{L}{ow} earth orbit (LEO) satellite communications will enhance next-generation wireless networks by providing global coverage and enabling ubiquitous connectivity \cite{lin2022path}.
With a large and ever-increasing number of LEO satellites deployed, satellite networks can provide robust communication services in areas where terrestrial deployment is insufficient, such as rural regions, mountains, oceans, and disaster-affected areas \cite{su2019broadband}.
In particular, satellite communications, and especially very small aperture terminal (VSAT)-class architectures with small satellite terminals, enable rapid deployment, cost-effective installation, and flexible scalability in remote and underserved areas. By leveraging these advantages, LEO satellite networks have great potential to be integrated with various mobile applications including backhaul systems \cite{wang2021joint}, mobile edge computing \cite{cao2022edge}, and integrated sensing and communication (ISAC) \cite{yin2024integrated}.

The focus for new spectrum in 6G networks is the mid-band spectrum, aka Frequency Range 3 (FR3), which refers to the 7–24 GHz range \cite{andrews20246,miao2023sub}, which significantly overlaps with spectrum allocated for satellite communications.  The lack of ``clean'' spectrum poses a major challenge to using the FR3 spectrum in 6G and highlights the importance of reliable and robust spectrum sharing between the satellite and terrestrial networks \cite{umer2025intelligent}. However, inter-system interference between the two systems severely degrades performance \cite{zhu2019cooperative}, especially with a denser deployment of base stations and satellite receivers. This is one of the greatest challenges in designing integrated satellite–terrestrial networks (ISTNs) \cite{peng2022integrating}.

Meanwhile, as the number of artificial intelligence (AI)-based services requiring rich contextual information related to the users’ location and environment increase, ISAC is also expected to become a key new functionality in 6G mobile networks \cite{liu2022integrated}.  Accordingly, in 6G, spectrum coexistence between satellite and ISAC-enabled terrestrial networks is highly likely \cite{lagunas20205g}, and therefore the impact of each system on the other should be carefully studied and optimized. 
Although the study in \cite{kim2024feasibility} primarily examines coexistence between different LEO satellite systems, its analysis on non-exclusive spectrum sharing and protection feasibility offers relevant insights for satellite–terrestrial coexistence scenarios.
The recent studies \cite{kong2025cooperative,pala2024empowering} have investigated satellite–terrestrial coexistence in ISAC systems, however, they mainly rely on ideal assumptions such as availability of perfect or statistical channel state information (CSI) and one-way interference from the satellite, leaving mutual satellite–terrestrial coupling largely unexplored.

\subsection{Background and Related Work}

\subsubsection{Beamforming Design for ISTN}
Several approaches have been proposed to enable spectrum sharing in ISTNs, primarily relying on terrestrial beamforming to suppress inter-system interference. Hybrid or analog beamforming has been studied to steer nulls toward satellite receivers or mitigate interference \cite{vazquez2016hybrid,peng2021hybrid,vazquez2018spectrum}. More advanced methods combine beamforming with resource allocation, such as joint scheduling, NOMA-based clustering, or time-slot assignment \cite{dong2023joint,lin2019joint}. Cognitive radio-based designs have also enforced interference constraints, either by protecting satellites via terrestrial beamforming or by optimizing satellite transmit power under terrestrial priority \cite{sharma2013transmit,ruan2019energy}.

While most prior studies on satellite–terrestrial integration have not explicitly considered beamforming for ISAC, a few recent works have attempted to address this problem. For example, \cite{kong2025cooperative} proposed a cooperative precoding method to balance communication and sensing, while \cite{pala2024empowering} applied federated learning to optimize beamforming and resource allocation with reconfigurable intelligent surfaces.

Several existing works consider the satellite as a passive entity \cite{vazquez2016hybrid,sharma2013transmit,vazquez2018spectrum}. Even when active functionalities are assumed, they typically rely on statistical CSI \cite{ruan2019energy,kong2025cooperative} or on idealized CSI acquisition \cite{ruan2019energy,peng2021hybrid,dong2023joint,lin2019joint}, which is infeasible in practice. Moreover, when designing satellite operation, many designs rely only on the satellite-to-terrestrial channel, neglecting CSI among terrestrial components, which results in performance limitation. Consequently, inter-system CSI sharing, which is essential for genuine co-design remains largely unaddressed.

\subsubsection{ISAC}
Based on receiver location, ISAC systems are classified into monostatic (co-located transmitter and receiver), bistatic (receiver is not near transmitter), and multistatic systems (multiple receivers).  In monostatic ISAC, since a terrestrial base station (TBS) serves as both transmitter and receiver, self-interference is a key challenge \cite{liu2023joint,he2023full}. While bistatic \cite{zhao2022joint} and multistatic systems \cite{lou2024beamforming,liu2024joint} are known to alleviate the limitations of monostatic radar, the increased number of receivers leads to additional deployment cost and higher complexity, and nontrivial coordination challenges.

As the number of sensor nodes in wireless networks increases, multistatic sensing, which leverages multiple sensors cooperatively to enhance sensing accuracy \cite{kaushik2024toward}, is expected to offer greater potential than single-node sensing operations \cite{liu2022integrated}.
While most existing works have focused on monostatic ISAC, the emergence of stringent sensing requirements envisioned for 6G, as discussed in\cite{ETSIGRISC001}, highlights the limitations of monostatic configurations in terms of coverage and full-duplex capability. Consequently, bistatic and multistatic ISAC have been highlighted in recent studies and reports as potential requirements for 6G \cite{ETSIGRISC001,gonzalez2024integrated}.
Accordingly, a comprehensive and practical study on multistatic ISAC is essential, particularly under realistic scenarios involving satellite interference.

Recently, several researchers have attempted to design ISAC techniques for satellite–terrestrial networks \cite{kong2025cooperative,pala2024empowering}. Although these methods jointly design beamforming at the satellite and terrestrial nodes, they assume that TBSs do not cause any interference to satellite terminals by disjointly partitioning coverage areas, which ultimately constrains the scalability of satellite terminal deployment.

As satellite networks and ISAC-enabled terrestrial networks are expected to share spectrum in 6G, it is important to explore their interactions. While prior works have investigated spectral coexistence, beamforming, and ISAC design, these studies remain largely independent. The integration of terrestrial ISAC with satellites is still underexplored, highlighting the need for unified approaches that jointly address co-design, inter-system interactions, and trade-offs.

\subsection{Our Contributions}

Our paper attempts to close the aforementioned gaps in the literature by providing a unified design for spectrum-sharing ISTNs, termed as \textit{integrated coexistence}. \textit{Integrated coexistence} refers to a coordinated design between satellite and terrestrial networks that enables feasible spectral coexistence through information sharing and mutual cooperation.
We highlight that the following practical considerations should be addressed.  

First, practical CSI sharing. It is infeasible to share real-time CSI between satellite and terrestrial networks, due to the large travel time from terrestrial networks to satellites \cite{wang2021joint}. Previous works have attempted to resolve this channel aging issue based on robust precoding \cite{wang2021joint,joroughi2019deploying,zhang2024joint} or AI-based CSI prediction methods \cite{abbasi2024channel,zhang2021deep,ying2024deep}. However, robust precoding methods suffer from performance limitations as worst-case optimization tends to produce overly conservative solutions, which can restrict achievable performance in practice. Meanwhile, AI-based CSI prediction relies on trained models, which imposes a heavy burden in acquiring datasets and training models. 

Second, practical beamforming.  When jointly optimizing satellite and terrestrial networks, a multi-beam satellite induces mutual coupling among numerous terrestrial cells within its footprint. A per-beam decomposition therefore cannot ensure optimality, while centralized coordination based on large-scale information exchange is challenging in practice.

Third, modeling all key sources of interference. From the perspective of terrestrial networks, it is necessary to avoid the strong interference originating from satellites and to protect satellite receivers from terrestrial interference. In addition, when considering the ISAC perspective, the sensing system must also be designed to mitigate satellite interference.

Our paper attempts to provide a practical solution for all three of those issues, in the context of multistatic ISAC, which is a generalization of bistatic ISAC and avoids full duplex receiver issues. The three main contributions of our paper are summarized as follows.

\textbf{Cooperative method design for ISTN.} Unlike prior designs where each network relies only on its local CSI to avoid interference, the proposed approach enables a more active co-design by sharing information between satellite and terrestrial nodes, and introduces a practical exchange method to realize cooperation. Specifically, since real-time CSI exchange is infeasible, the method exploits the predictability of satellite CSI by pre-optimizing beamforming and power allocation in advance and subsequently refining beamforming with terrestrial CSI. Furthermore, instead of a centralized approach requiring extensive information exchange across multiple terrestrial cells, the proposed method relies on distributed optimization, thereby reducing coordination overhead.

\textbf{Beamforming and satellite power allocation method design for multistatic ISAC.} A co-design of terrestrial beamforming under satellite interference with a satellite spot-beam power allocation preventing excessive interference toward terrestrial network is proposed to maximize the sum rate of terrestrial and satellite networks. First, the applicability of the signal-to-clutter-plus-noise ratio (SCNR) metric to the weighted minimum mean-squared-error (WMMSE) algorithm is examined. Next, a WMMSE-based sum rate maximization algorithm for ISTNs is designed that ensures both minimum sensing and satellite rate requirements. In addition, we provide a target–radar receiver association method considering satellite and inter-target interference.  
Unlike conventional interference-constrained methods relying solely on satellite CSI or treating satellites as passive entities, the proposed method actively co-designs both networks by leveraging cross-system information, thereby guaranteeing sensing and communication performance.

\textbf{Feasibility study on spectral coexistence of satellite and terrestrial networks with ISAC.} Through simulations, the effectiveness of the proposed method and the feasibility of spectral coexistence are examined. In particular, the impact of LEO satellite interference on terrestrial sensing performance is studied. 
Through simulation, it is shown that increasing the number of spot beams and radar antennas reduces the performance gap between the interference-free scenario and the proposed method, demonstrating the possibility of spectral coexistence between the two systems without significant degradation.

\section{System Model}

\begin{figure}[t]
    \centerline{\includegraphics[trim={2.8cm 1.8cm 12cm 2.75cm},clip,width=0.94\linewidth]{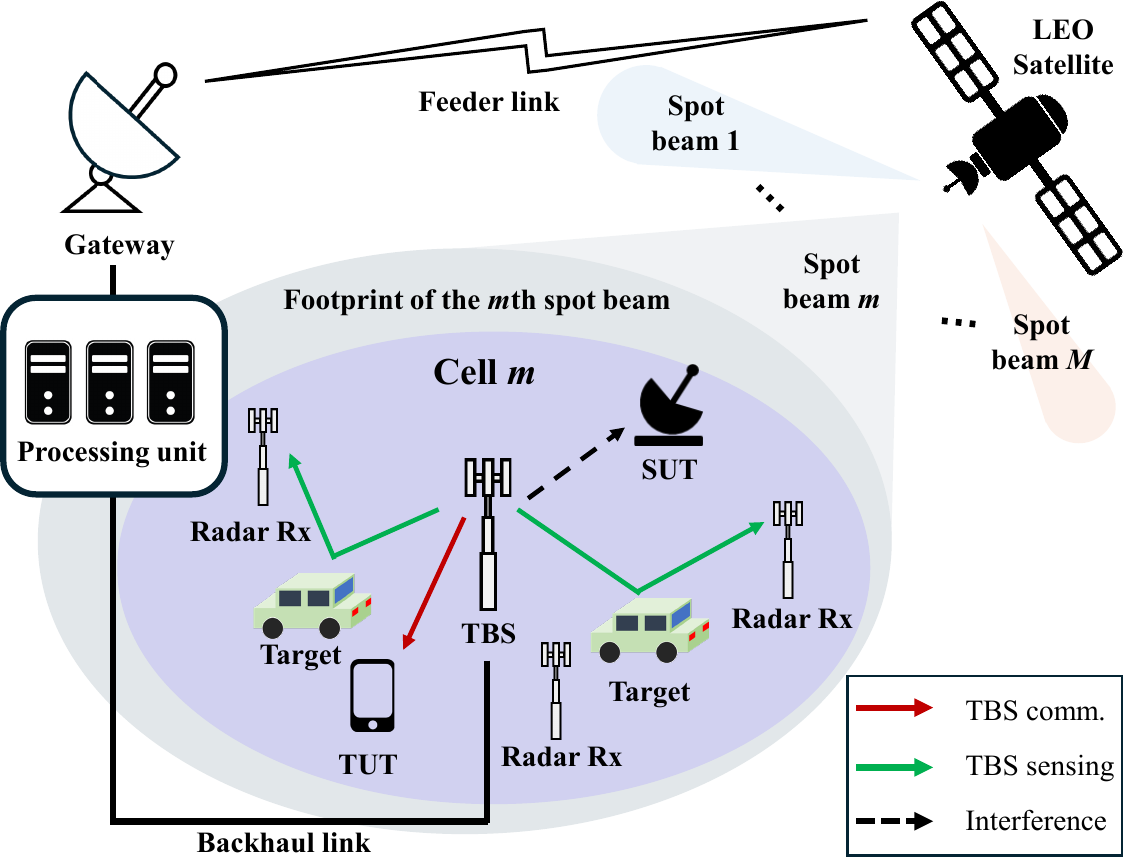}}
    \caption{An illustration of system model. The LEO satellite covers multiple areas with multiple spot beams. The coverage of the $m$th terrestrial cell is included in the footprint of the $m$th spot beam of LEO satellite, generating interference. In each terrestrial cell, a TBS transmits ISAC signal toward multiple TUTs and targets, causing interference toward nearby SUTs. 
    The echo signal reflected from each target is acquired by radar receivers for sensing. The TBS is connected with a gateway equipped with a processing unit, which enables cooperation between two networks.}
    \label{FIG:sys_model} 
\end{figure}
As shown in Fig. \ref{FIG:sys_model}, a downlink ISTN with a multi-beam LEO satellite and a terrestrial network that share the same frequency spectrum is considered. As a worst-case assumption, the terrestrial and satellite networks are assumed to transmit concurrently with complete time–frequency overlap. The LEO satellite operates $M$ spot beams simultaneously to serve $M$ designated satellite user terminals (SUTs), which are assumed to be VSAT-class terminals with single small-aperture directional antennas, and we focus on $M$ terrestrial cells located within the coverage of each spot beam\footnotemark. 
\footnotetext{Although a single terrestrial cell per spot beam is considered for simplicity of the system model and notations, extending this work to systems with multiple terrestrial cells per spot beam is straightforward, which will be given in the next section.}
In this paper, the terrestrial cell affected by the $m$th spot beam is denoted as the $m$th cell. Typically, the coverage of a spot beam is larger than terrestrial cells, and it is assumed that the coverage area the $m$th terrestrial cell is fully included in the $m$th spot beam footprint.
With a sufficiently large distance among spot beams, inter-cell interference among terrestrial networks is assumed to be negligible.
The satellite cooperates with the terrestrial network through gateways and feeder links, where each gateway is connected to a processing unit that performs centralized computations and forwards optimized spot beam power allocation commands. The backhaul and feeder links are assumed to be lossless.

The ISAC-enabled terrestrial networks support communication users while simultaneously transmitting radar signals for sensing targets. Each terrestrial cell consists of one TBS with $N_{\rm TX}$ antennas, $K$ terrestrial user terminals (TUTs) each with a single antenna, $N_{\rm tar}$ sensing targets, and $N_{\rm rad}$ radar receivers, each equipped with $N_{\rm RX}$ receive antennas. The array configurations of both the TBS and the radar receivers are assumed to be uniform linear arrays (ULAs). The TBS transmits communication data streams toward the $K$ TUTs while simultaneously transmitting radar signals to the $N_{\rm tar}$ sensing targets. Each radar receiver obtains echoes reflected from the targets assigned to it according to a predefined association rule. 
To reduce estimation complexity, each target is assumed to be associated with a single radar receiver.

\subsection{Channel Model}
The attenuation coefficient between the LEO satellite and terrestrial components is expressed as follows \cite{choi2023joint}.
\begin{align}
    g_{m,k}^{\rm X} = \left(\frac{c}{2 \pi f_c} \right)^2 \frac{G_s G_m^{\rm X}}{(d_{m,k}^{\rm X})^2} B_{m,k}^{\rm X} \xi,
\end{align}
where ${\rm X} \in \{\rm SUT,TUT,rad \}$, $f_c,c,d_{m,k}^{\rm X},G_s, G_m^{\rm X},B_{m,k}^{\rm X}$ and $\xi$ are defined as carrier frequency, the speed of light, distance between ground nodes and satellite, antenna gain for LEO satellite, receive antenna gain of ground nodes, beam pattern for LEO satellite, and shadowing factor, respectively. 
The receive antenna gain for an SUT is modeled as a rectangular antenna pattern \cite{dabiri20203d} defined as
\[
    G_m^{\rm SUT} =
\begin{cases}
    G_{\rm main}, & \text{if } |\phi| \le \phi_{\rm th},\\
    G_{\rm side}, & \text{otherwise},
\end{cases}
\]
where $G_{\rm main}$ and $G_{\rm side}$ are the main-lobe and side-lobe gains, $\phi$ is the elevation angle offset with respect to the satellite direction, and $\phi_{\rm th}$ is its threshold.  
The beam pattern toward terrestrial components $B_{m,k}^{\rm X}$ is modeled as \cite{choi2023joint}
\begin{align}
    B_{m,k}^{\rm X} = 10^{-0.3 \left(\frac{\varphi_{m,k}^{\rm X}}{\varphi_{\rm 3dB}} \right)^2 },
\end{align}
where $\varphi_{m,k}^{\rm X}$ and $\varphi_{\rm 3dB}$ denote the angular separation measured from the center of the $k$th beam to the $m$th component, and the 3 dB angle, respectively. The $m$th spot beam is assumed to be centered on the $m$th SUT. The shadowing factor $\xi$ follows a log-normal distribution with standard deviation $\sigma_{\xi}$ \cite{choi2023joint}.

Similarly, the channel between the LEO satellite and the $n$th radar receiver in the $m$th cell is modeled as follows.
\begin{align}
    {\bf g}_{m,n}^{\rm rad} = \sqrt{g_{m,n}^{\rm rad}} {\bf a}_{\rm r} (\theta_{m,n}^{\rm sat}),
\end{align}
where ${\bf a}_{\rm r}(\cdot)$ is the receive array response vector for the ULA, and $\theta_{m,n}^{\rm sat}$ is the angle-of-arrival (AOA) of the satellite with respect to the $n$th radar receiver in the $m$th cell.

In the $m$th terrestrial cell, the communication channel between the TBS and the $k$th TUT is represented as
\begin{align}
    {\bf h}_{m,k} = P^{\rm LOS}_{m,k} \sqrt{\beta_{m,k}^{(\rm L)}} {\bf h}_{m,k}^{(\rm L)} + \sqrt{\beta_{m,k}^{(\rm N)}} {\bf h}_{m,k}^{(\rm N)},
\end{align}
where $P^{\rm LOS}_{m,k}$ indicates the probability of a line-of-sight (LOS) connection. $\beta_{m,k}^{(\rm L)}$ and $\beta_{m,k}^{(\rm N)}$ denote the large-scale fading components corresponding to LOS and non-line-of-sight (NLOS) components, respectively. The small-scale fading components are modeled as
\begin{align}
    {\bf h}_{m,k}^{(\rm L)} &= \sqrt{N_{\rm TX}} \alpha_k^{(\rm L)} {\bf a}_{\rm t}^{\rm H} \left(\theta_k^{(\rm L)}\right),\\
    {\bf h}_{m,k}^{(\rm N)} &= \sqrt{\frac{N_{\rm TX}}{N_{\rm cluster}N_{\rm ray}}} \sum_{i=1}^{N_{\rm cluster}} \sum_{j=1}^{N_{\rm ray}} \alpha_{k,i,j}^{(\rm N)} {\bf a}_{\rm t}^{\rm H} \left(\theta_{k,i,j}^{(\rm N)}\right),
\end{align}
where ${\bf a}_{\rm t} (\cdot)$ denotes the transmit array response vector. The parameters $\alpha_k^{(\rm L)}$ and $\alpha_{k,i,j}^{(\rm N)}$ represent the complex path gains for LOS and NLOS components, respectively. $\theta_k^{(\rm L)}$ is the angle-of-departure (AOD) of the LOS path, while $\theta_{k,i,j}^{(\rm N)}$ denotes the AOD of the NLOS rays. $N_{\rm cluster}$ and $N_{\rm ray}$ are the number of clusters per channel and the number of rays per cluster, respectively.  
Similarly, the propagation channel between the TBS and an SUT is modeled in the same way as the terrestrial channel, i.e.,
\begin{align}
    {\bf h}_{m}^{(\rm S)} = P^{\rm LOS}_{m,(\rm S)} \sqrt{\gamma_m^{(\rm L)}} {\bf h}_{m,(\rm S)}^{(\rm L)} + \sqrt{\gamma_m^{(\rm N)}} {\bf h}_{m,(\rm S)}^{(\rm N)} ,
\end{align}
where $\gamma_m^{(\rm L)} = G_{m}^{\rm SUT} \beta_{m,(\rm S)}^{(\rm L)}$ and $\gamma_m^{(\rm N)} = G_{m}^{\rm SUT} \beta_{m,(\rm S)}^{(\rm N)}$. The parameters are assumed to have the same statistics as those of the terrestrial channel.

Although a number of works assume a single-path LoS sensing channel, real environments with multistatic ISAC can produce additional echoes from unintended targets and clutter. This work addresses these complexities for a more accurate sensing system model.
The sensing channel between the TBS and the $n$th radar receiver inside the $m$th cell is expressed as
\begin{align}
    {\bf G}_{m,n} = \sum_{i=1}^{N_{\rm tar}} \alpha_{i,n,m}^{\rm tar} {\bf a}_{\rm r} (\theta_{i,n,m}^{\rm r}) {\bf a}_{\rm t} (\theta_{i,m}^{\rm t}) = \sum_{i=1}^{N_{\rm tar}} {\bf G}_{i,n,m}^{\rm tar},
\end{align}
where $\theta_{i,n,m}^{\rm r}$ and $\theta_{i,m}^{\rm t}$ denote the angle from the $i$th target to the $n$th receiver, and the angle from the TBS to the $i$th target in the $m$th cell, respectively. To ensure that the sensing operation is feasible, it is assumed that a LoS path between the radar receiver and the target always exists. $\alpha_{i,n,m}^{\rm tar}$ denotes the path gain from the $i$th target to the $n$th receiver in the $m$th cell, which includes path-loss and radar cross-section (RCS) gain, and is modeled as \cite{yu2025multi}
\begin{align}
    \alpha_{i,n,m}^{\rm tar} = \frac{\lambda^2 \sigma_{i,n,m}^2}{(4\pi)^3 (d_{n,m}^{\rm tx})^2 (d_{i,n,m}^{\rm rx})^2},
\end{align}
where $\lambda, \sigma_{i,n,m}, d_{n,m}^{\rm tx}$, and $d_{i,n,m}^{\rm rx}$ are the wavelength, RCS gain, distance from the $m$th TBS to the $i$th target, and distance from the $i$th target to the $n$th receiver in the $m$th cell, respectively.  

Furthermore, the presence of clutter around the sensing target generates additional echoes, which can degrade sensing performance. The propagation channel from the $l$th clutter to the $n$th radar receiver in the $m$th cell is expressed as follows.
\begin{align}
    {\bf G}_{l,n,m}^{\rm cl} = \alpha_{l,n,m}^{\rm cl} {\bf a}_{\rm r} (\theta_{l,n,m}^{\rm r,cl}) {\bf a}_{\rm t} (\theta_{l,m}^{\rm t,cl}),
\end{align}
where $\alpha_{l,n,m}^{\rm cl}$ is the path gain from the $l$th clutter to the $n$th radar receiver, and $\theta_{l,n,m}^{\rm r,cl}$ and $\theta_{l,m}^{\rm t,cl}$ denote the angle from the $l$th clutter to the $n$th receiver and the angle from the TBS to the $l$th clutter in the $m$th cell, respectively.

\subsection{Signal Model}

Each ISAC-enabled TBS simultaneously transmits signals dedicated to communication and sensing. The signal vector for the $m$th TBS is defined as ${\bf s}_m = [s_{m,1},\cdots,s_{m,K+N_{\rm tar}}]^{\rm T}$, which contains the data stream for the $k$th TUT in $s_{m,k}$ and the radar signal for the $i$th target in $s_{m,K+i}$, and satisfies ${\bf s}_m \sim \mathcal{CN}({\bf 0},{\bf I}_{K+N_{\rm tar}})$. The $m$th TBS transmits the ISAC signal via the beamforming matrix ${\bf F}_m = [{\bf f}_{m,1},\cdots,{\bf f}_{m,K+N_{\rm tar}}]$, where ${\bf f}_{m,k}$ and ${\bf f}_{m,K+i}$ denote the beamforming vectors for the $k$th TUT and the $i$th target, respectively.

From the channel model described above, the received signal for the $k$th TUT in the $m$th cell can be expressed as
\begin{align}
    y_{m,k} &= {\bf h}_{m,k}{\bf f}_{m,k} s_{m,k} + \sum_{j=1, j\neq k}^{K+N_{\rm tar}} {\bf h}_{m,k}{\bf f}_{m,j} s_{m,j}\nonumber \\
    & \,\,+ \sqrt{g_{m,k}^{\rm TUT} p_m^{(\rm S)}} s_m^{(\rm S)} + n_{m,k},
\end{align}
where $p_m^{(\rm S)}$ denotes the transmit power allocated to the $m$th spot beam of the LEO satellite and $s_m^{(\rm S)}$ is the data stream for the $m$th SUT. $n_{m,k}$ is additive white Gaussian noise (AWGN) with zero-mean and variance $\sigma_{\rm n,T}^2$.

The received signal for the $m$th SUT is given by
\begin{align}
    y_m^{(\rm S)} = \sqrt{g_{m}^{\rm SUT} p_m^{(\rm S)}} s_m^{(\rm S)} + \sum_{j=1}^{K+N_{\rm tar}} {\bf h}_{m}^{(\rm S)}{\bf f}_{m,j} s_{m,j} + n_m^{(\rm S)},
\end{align}
where $n_{m}^{(\rm S)}$ is AWGN at the $m$th satellite receiver with zero-mean and variance $\sigma_{\rm n,S}^2$. 

The received signal for the $n$th radar receiver is modeled as
\begin{align}
    {\bf y}_{m,n}^{\rm rad} = \bar {\bf G}_{m,n} {\bf F}_m {\bf s}_m + {\bf g}_{m,n}^{\rm rad} \sqrt{p_m^{(\rm S)}} s_m^{(\rm S)}  + {\bf n}_{m,n}^{\rm rad}, 
\end{align}
where $\bar {\bf G}_{m,n} = {\bf G}_{m,n}+ {\bf G}_{m,n}^{\rm cl}$, ${\bf G}_{m,n}^{\rm cl} = \sum_{l} {\bf G}_{l,m,n}^{\rm cl}$, and ${\bf n}_{m,n}^{\rm rad}$ is the AWGN vector for the $n$th radar receiver in the $m$th cell with ${\bf n}_{m,n}^{\rm rad} \sim \mathcal{CN}({\bf 0},\sigma_{\rm n,R}^2 {\bf I}_{N_{\rm RX}})$. To improve sensing performance, each radar receiver employs receive beamforming to suppress interference. Let ${\bf w}_{i,m}$ denote the beamforming vector to receive the echo from the $i$th target at the $A(i,m)$th receiver in the $m$th cell, where the association operator $A(i,m)$ is defined as
\begin{align}
A:\{(i,m)\mid m\!\in\!\mathcal{M},~ i\!\in\!\mathcal{T}_m\}\!\to\!\mathcal{R}_m,\,\,n_{i,m} = A(i,m),
\end{align}
with $\mathcal{M}$, $\mathcal{T}_m$, $\mathcal{R}_m$, and $n_{i,m}$ denoting the set of cells, the targets in the $m$th cell, the radar receivers in the $m$th cell, and the index of the radar receiver associated with the $i$th target in the $m$th cell, respectively.  
The received signal for detecting the echo from the $i$th target, $r_{i,m} = {\bf w}_{i,m}{\bf y}_{m,A(i,m)}^{\rm rad}$ becomes
\begin{align}
    r_{i,m} &= {\bf w}_{i,m} \Bigl( {\bf G}_{i,A(i,m),m}^{\rm tar} {\bf f}_{m,K+i} s_{m,K+i} \nonumber\\
    &+ \sum_{j=1, j\neq i}^{N_{\rm tar}} {\bf G}_{j,A(i,m),m}^{\rm tar} {\bf f}_{m,K+i} s_{m,K+i} \nonumber\\
    &+ {\bf G}_{m,A(i,m)}^{\rm cl} {\bf F}_{m} {\bf s}_{m}+ \sum_{j=1, j\neq K+i}^{K+N_{\rm tar}} {\bf G}_{m,A(i,m)} {\bf f}_{m,j} s_{m,j} \nonumber\\
    &+ {\bf g}_{m,A(i,m)}^{\rm rad} \sqrt{p_m^{(\rm S)}} s_m^{(\rm S)} 
    + {\bf n}_{m,A(i,m)}^{\rm rad} \Bigr).
\end{align}
In the equation above, each term corresponds to the desired echo signal, interference from unintended targets, interference from clutter, interference from other beamforming vectors, interference from the LEO satellite, and noise, respectively. 
Unlike monostatic sensing systems with single target, echoes from unintended targets and clutter generate additional interference from undesired directions, which distracts sensing operation although they are coherent with the desired echo.

From the definitions above, the signal-to-interference-plus-noise ratio (SINR) of the TUTs is defined as
\begin{align}
    {\rm SINR}_{m,k} = \frac{{|{\bf h}_{m,k}{\bf f}_{m,k}|}^2}{\sum_{j\neq k} {|{\bf h}_{m,k}{\bf f}_{m,j}|}^2 + g_{m,k}^{\rm TUT} p_m^{(\rm S)} + \sigma_{\rm n,T}^2 }.
\end{align}
Likewise, the SINR for the $m$th SUT is
\begin{align}
    {\rm SINR}_{m}^{(\rm S)} = \frac{g_{m}^{\rm SUT} p_m^{(\rm S)}}{\sum_{j=1}^{K+N_{\rm tar}} {|{\bf h}_{m}^{(\rm S)}{\bf f}_{m,j}|}^2 + \sigma_{\rm n,S}^2 }.
\end{align}
For wireless sensing, the SCNR \cite{choi2024joint} is adopted as the performance metric. The SCNR for the $i$th target in the $m$th cell is defined as 
\begin{align}
    {\rm SCNR}_{i,m} = \frac{{|{\bf w}_{i,m} {\bf G}_{i,A(i,m),m}^{\rm tar} {\bf f}_{m,K+i}|}^2}{\sum_{k=1}^4 I_{m,i}^{(k)} + \|{\bf w}_{i,m}\|^2 \sigma_{\rm n,R}^2}, 
\end{align}
where 
\begin{align}
    I_{m,i}^{(\rm 1)} &= \sum_{j=1, j\neq i}^{N_{\rm tar}} \left|{\bf w}_{i,m}  {\bf G}_{j,A(i,m),m}^{\rm tar} {\bf f}_{m,K+i}\right|^2, \\I_{m,i}^{(\rm 2)} &= \sum_l \|{\bf w}_{i,m} {\bf G}_{l,A(i,m),m}^{\rm cl} {\bf F}_{m}\|^2,\\ 
    I_{m,i}^{(\rm 3)} &= \sum_{j\neq K+i} \left|{\bf w}_{i,m} {\bf G}_{m,A(i,m)} {\bf f}_{m,j}\right|^2,\\ 
    I_{m,i}^{(\rm 4)} &= \left|{\bf w}_{i,m} {\bf g}_{m,A(i,m)}^{\rm rad}\right|^2 p_m^{(\rm S)}.
\end{align}

Finally, the achievable rate for the $k$th TUT in the $m$th cell and the $m$th SUT are defined as
\begin{align}
    R_{m,k}^{(\rm T)} &= \log_2 \left(1 + {\rm SINR}_{m,k}\right),\\
    R_m^{(\rm S)} &= \log_2 \left(1 + {\rm SINR}_{m}^{(\rm S)}\right).
\end{align}

\noindent\textbf{Main assumptions.} The main assumptions in this work are given as follows.
(i) Satellite CSI can be reliably predicted for a short time interval. 
Here, the time interval is defined as the duration sufficient to deliver the required information from the terrestrial side to the satellite through the gateway. 
(ii) During the aforementioned latency interval, the large-scale components (path-loss and blockage) of the terrestrial and satellite channels remain approximately constant without abrupt variations.
(iii) Information from terrestrial network can be shared reliably via lossless backhaul and feeder link.

\section{Proposed ISTN Design}
In this section, the cooperative design for ISTNs with multistatic ISAC systems is proposed. The achievable sum rate maximization problem subject to the minimum rate constraint for SUTs and the minimum SCNR constraint for sensing operations is formulated. The design jointly considers the ISAC beamforming matrix at each TBS and the power allocation for each satellite spot beam.
\begin{subequations}\label{problem1}
\begin{align}
&\mathop {\rm maximize}\limits_{\{{\bf F}_m\},A,\{{\bf w}_{i,m}\},\{p_m^{(\rm S)}\}} \,\,\,\, \sum_{m=1}^M \left(\sum_{k=1}^K R_{m,k}^{(\rm T)} + R_m^{(\rm S)} \right) \tag{\theequation a}\label{cost}\\
&\quad\quad\,\mathop {\rm subject\; to}\qquad\,\,\, R_m^{(\rm S)} \ge R_{\rm min}^{(\rm S)}, \forall m, \label{constraint1} \tag{\theequation b}\\
&\quad\quad\qquad\qquad\qquad\quad {\rm SCNR}_{i,m} \ge {\rm SCNR}_{\rm min}, \forall i,m, \tag{\theequation c} \label{constraint2}\\
&\quad\quad\qquad\qquad\qquad\quad {\rm tr}\left({\bf F}_m^{\rm H} {\bf F}_m \right) \le P_{{\rm BS},m}, \forall m, \tag{\theequation d} \label{constraint3}\\
&\quad\quad\qquad\qquad\qquad\quad \sum_{m=1}^M p_m^{(\rm S)} \le P_{\rm LEO}. \tag{\theequation e} \label{constraint4}
\end{align}
\end{subequations}
The constraints \eqref{constraint1}–\eqref{constraint4} represent the minimum rate constraint for SUTs, the minimum SCNR constraint for sensing, the power constraints of the TBS and the LEO satellite, respectively.

This formulation results in a nonconvex optimization problem combined with a combinatorial assignment problem, leading to highly complex binary integer programming. Furthermore, due to the transmission latency between terrestrial networks and satellites, it is challenging to fully exploit instantaneous information such as CSI in real time \cite{wang2021joint,arnau2014performance}. The transmitted CSI or power allocation optimized based on CSI is typically outdated due to channel aging, making full cooperation based on instantaneous information extremely difficult. 
In addition, multiple terrestrial cells are affected by the multiple spot beams, which implies mutual coupling among satellite power allocation for each SUT. This necessitates a centralized optimization with global information sharing, which requires significant latency.

To address this problem, the following approach is proposed. First, we design a cooperation structure between the satellite and terrestrial networks to circumvent the channel aging issue and to enable joint optimization in a distributed manner across multiple cells. Then, beamforming and satellite power allocation techniques are developed to mitigate inter-system interference.

\subsection{Proposed Cooperation Structure}

\begin{figure}[h]
    \centerline{\includegraphics[trim={8.7cm 1.7cm 8cm 2.3cm},clip,width=0.94\linewidth]{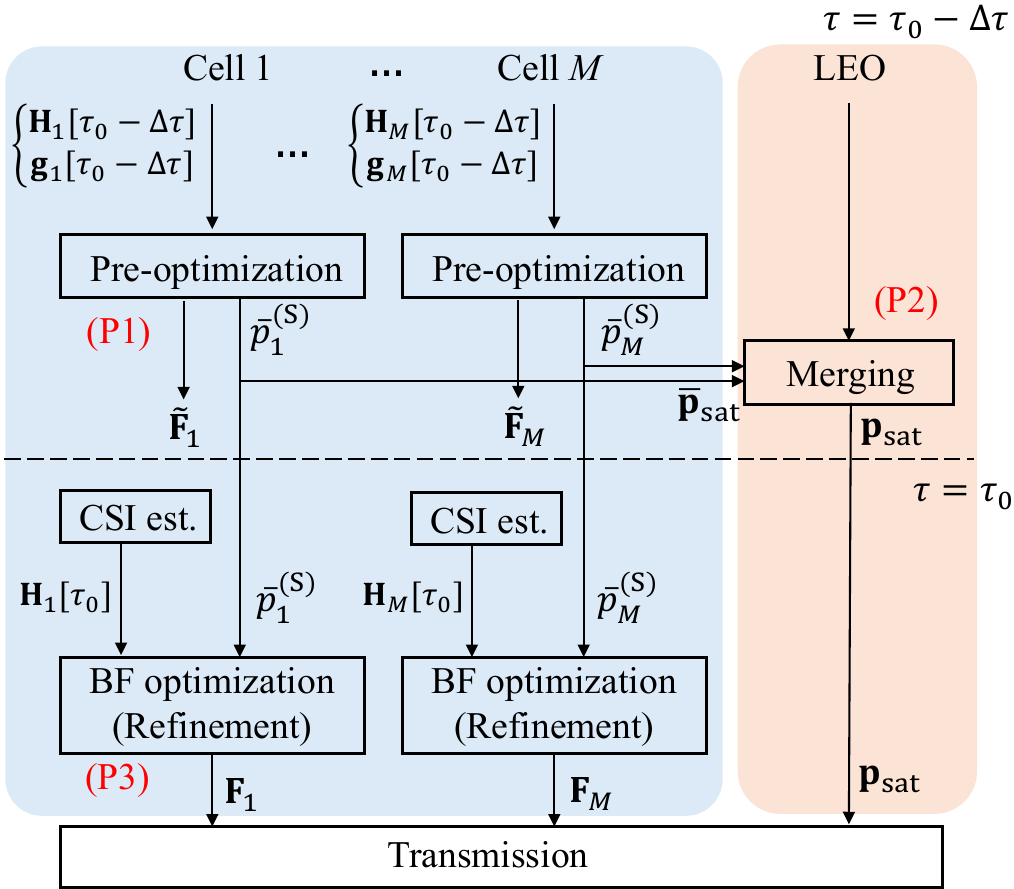}}
    \caption{Illustration of the proposed cooperation structure. For simplicity, the terrestrial CSI and satellite CSI for the $m$th cell at time slot $\tau$ are expressed as ${\bf H}_m[\tau]$ and ${\bf g}_m[\tau]$, respectively.}
    \label{FIG:cooperation} 
\end{figure}

We propose a cooperative structure illustrated in Fig.~\ref{FIG:cooperation}. The overall process consists of three stages, pre-optimization (P1), merging (P2), and refinement (P3).

The proposed structure is based on the observation that satellite CSI is more predictable than terrestrial CSI. This is because satellite CSI is LOS-dominant and mainly determined by path-loss and direction, whereas terrestrial CSI consists of many spatial paths with complex reflection characteristics, making it highly sensitive to abrupt changes caused by the movement of nearby objects. In addition, since satellite CSI is primarily determined by fixed orbital trajectories, it can be predicted more easily. On the other hand, terrestrial CSI varies rapidly with larger uncertainties. 
The latency of the terrestrial-to-satellite link is on the order of several tens of milliseconds \cite{abderrahim2020latency}, during which only the small-scale channel coefficients experience significant variation, while large-scale effects such as shadowing, blockage, and path-loss remain relatively constant.

Exploiting this property, the proposed method is designed under the assumption that satellite CSI can be reliably predicted for a short time interval. Specifically, let us assume that satellite power and terrestrial beamforming are designed for $\tau = \tau_0$. The pre-optimization stage is performed at $\tau = \tau_0 - \Delta\tau$, where $\Delta\tau$ is the time required for information transmission from the terrestrial network to the satellite.

At $\tau = \tau_0 - \Delta\tau$, each terrestrial network predicts the satellite’s trajectory at $\tau = \tau_0$ and optimizes the joint terrestrial beamforming and satellite power allocation in advance using the current terrestrial CSI and predicted satellite CSI (P1). This is feasible since satellite power allocation is primarily affected by the large-scale attenuation of the terrestrial CSI, and the small-scale perturbations have negligible impact on the overall optimization result. The optimized power allocation vector is then sent to the satellite network, specifying how the power should be distributed at $\tau = \tau_0$. 

However, since the power is optimized in a distributed manner, without incorporating information from other cells, the results may not satisfy constraint \eqref{constraint4}. To address this, a merging process (P2) is introduced to centrally adjust the results, thereby ensuring \eqref{constraint4} and enabling network-wide optimization that yields additional performance gains. Finally, at $\tau = \tau_0$, the satellite allocates transmit power according to the pre-determined vector from (P2), while the terrestrial network, which has prior knowledge of the expected interference, estimates terrestrial CSI and refines the beamforming matrix in a refinement stage (P3).

\subsection{Pre-optimization Stage}
\label{Sec:singlecell} 
First, the pre-optimization in the $m$th cell is expressed as
\begin{equation}\label{P1}
\begin{aligned}
{\rm P1} :\quad &\mathop {\rm maximize}\limits_{{\bf F}_m,A,{\bf w}_{i,m},p_m^{(\rm S)}} \,\,\,\, \sum_{k=1}^K R_{m,k}^{(\rm T)} + R_m^{(\rm S)} \\
&\,\,\,\,\mathop {\rm subject\; to}\,\,\,\, R_m^{(\rm S)} \ge R_{\rm min}^{(\rm S)}, \\
&\,\,\,\,\qquad\qquad\quad {\rm SCNR}_{i,m} \ge {\rm SCNR}_{\rm min}, \forall i, \\
&\,\,\,\,\qquad\qquad\quad {\rm tr}\left({\bf F}_m^{\rm H} {\bf F}_m \right) \le P_{{\rm BS},m}, \\
&\,\,\,\,\qquad\qquad\quad p_m^{(\rm S)} \le P_{\rm LEO}.
\end{aligned}
\end{equation}
Note that \eqref{P1} is a distributed single-cell optimization problem in which the sum-power constraint across all spot beams is replaced with a per-spot beam power constraint.
Unlike conventional monostatic ISAC systems, target-receiver association should be designed since LEO satellite induces directional interference toward radar receivers, which distracts sensing operation.
To solve this problem, we divide \eqref{P1} into two subproblems, target–receiver association $A$ and beamforming ${\bf F}_m, {\bf w}_{i,m}$ with power allocation $p_m^{(\rm S)}$.

\subsubsection{Target-Receiver Association}
The objective of the target–receiver association is to assign each target to the most suitable radar receiver so that the receiver can capture the target’s echo without being severely disturbed by interference. The association is performed in such a way that the receiver obtaining the target’s echo is not aligned with interference from other targets in similar directions. In addition, within an ISTN, signals from LEO satellites may act as a source of interference. Therefore, the AOA of satellite interference must also be taken into account for radar receivers. 

The target-receiver association algorithm is expressed as follows.
First, the initial candidate set $\mathcal{I}_m$ is defined as the set of all possible target–receiver index pairs in the $m$th cell. If the angular difference between the AOA of a target–receiver pair and that of the corresponding satellite–receiver pair is smaller than $\delta_{\text{sat}}$, the pair is discarded from $\mathcal{I}_m$. Subsequently, targets are matched to radar receivers in descending order of path gain $\alpha_{i,n,m}^{\rm tar}$ since echoes with larger path gain are more advantageous for estimation. After a target–receiver pair is matched, any other targets whose angular difference from the matched target with respect to the same receiver is smaller than $\delta_{\text{tar}}$ are excluded from $\mathcal{I}_m$. The algorithm terminates once every target has been matched to exactly one receiver.
The proposed association algorithm can be summarized as Algorithm \ref{alg1}.

\begin{algorithm}[t]
\caption{Target--Receiver Association Algorithm}
\label{alg1}
\begin{algorithmic}[1]
\Require $\{\theta_{i,n,m}^{\rm r}\}$, $\{\theta_{m,n}^{\rm sat}\}$, $\{\alpha_{i,n,m}^{\rm tar}\}$, $\delta_{\text{sat}}$, $\delta_{\text{tar}}$
\State $\mathcal{I}_m \gets \{(i,n)\mid i\in\{1,\dots,N_{\rm tar}\},\; n\in\{1,\dots,N_{\rm rad}\}\}$
\For{each cell $m$}
    \For{each $(i,n)\in \mathcal{I}_m$}
        \If{$|\theta_{i,n,m}^{\rm r}-\theta_{m,n}^{\rm sat}| < \delta_{\text{sat}}$}
            \State $\mathcal{I}_m \gets \mathcal{I}_m \setminus \{(i,n)\}$
        \EndIf
    \EndFor
    \State Sort $(i,n)\in\mathcal{I}_m$ by $\alpha_{i,n,m}^{\rm tar}$ in descending order
    \For{each target $i$}
        \If{$i$ not assigned}
            \State $n^{*} \gets \arg\max_{n:(i,n)\in\mathcal{I}_m} \alpha_{i,n,m}^{\rm tar}$
            \State $\mathcal{I}_m \gets \mathcal{I}_m \setminus \{(i,n): n\neq n^{*}\}$
            \State $\mathcal{I}_m \gets \mathcal{I}_m \setminus \{(i',n^{*}): i'\neq i,~|\theta_{i,n^{*},m}^{\rm r}-\theta_{i',n^{*},m}^{\rm r}| < \delta_{\text{tar}}\}$
        \EndIf
        \If{all targets assigned}
            \State \textbf{break}
        \EndIf
    \EndFor
    \For{each target $i$}
        \State $A(i,m) \gets \{\,n \mid (i,n)\in\mathcal{I}_m\,\}$
    \EndFor
\EndFor
\end{algorithmic}
\end{algorithm}

\subsubsection{Joint Beamforming and Power Allocation Design}
We now design a beamforming and power allocation method for single-cell ISTNs. It is assumed that the target–receiver association is predefined using Algorithm \ref{alg1}. Unlike several prior works \cite{vazquez2016hybrid,vazquez2018spectrum,sharma2013transmit} in which each network relies solely on its own CSI to mitigate mutual interference, this work designs a cooperation method which directly guarantees the performance requirements of both networks utilizing predicted satellite CSI and current terrestrial CSI.

To solve \eqref{P1}, the achievable sum rate of the ISTN should be maximized subject to minimum performance constraints for the SUT and the SCNR. If we regard the satellite and SUT as a TBS and a TUT, respectively, Problem \eqref{P1} becomes similar to a beamforming problem for a multi-input multi-output (MIMO) interfering broadcast channel with multiple TBSs \cite{shi2011iteratively,padmanabhan2016precoder,ma2016qos}, where quality-of-service (QoS)-constrained beamforming design is possible using modified WMMSE algorithms \cite{padmanabhan2016precoder,ma2016qos}. 

Our proposed methodology incorporates both satellite and sensing systems into the WMMSE framework \cite{shi2011iteratively}. 
However, undesired reflections from other targets and clutters, although coherent with the desired echo, act as interference that degrades sensing performance. Therefore, within the SCNR framework, the well-known relationship between SINR and the minimum mean-squared-error (MMSE), $1 + {\rm SINR} = {\rm MMSE}^{-1}$, does not hold in general. Since the WMMSE algorithm relies on this identity to transform the rate expression into a mean-squared-error (MSE) formulation, its direct application to SCNR is not straightforward.

In this work, the applicability of the WMMSE algorithm for ISAC systems is demonstrated with the following example. For simplicity, the received signal for radar and the corresponding SCNR can be defined as
\begin{align}
    &y_s = \mathbf{wH}_d \mathbf{f}_d s_d + \mathbf{wH}_c \mathbf{f}_d s_d+ \sum_{j=1}^L \mathbf{wH}_j \mathbf{f}_j s_j + {\bf wn}, \\
    &\text{SCNR} = \frac{|\mathbf{wH}_d \mathbf{f}_d|^2}{|\mathbf{wH}_c \mathbf{f}_d|^2 + \sum_{j=1}^L |\mathbf{wH}_j \mathbf{f}_j|^2 + \sigma^2 \|{\bf w}\|^2},
    \label{eq:scnr}
\end{align}
where $\mathbf{H}_d, \mathbf{H}_c$, $\mathbf{H}_j$, and $\bf w$ denote the channel matrices for the desired target, clutter, the $j$th target/TUT, and receive beamforming vector, respectively. Similarly, $\mathbf{f}_d$ and $\mathbf{f}_j$
are the beamforming vectors for the desired target and the $j$th target/TUT, respectively. 

Next, the received signal of a virtual communication system is defined by converting coherent echo component into non-coherent interference.
\begin{align}
    y_c &= \mathbf{wH}_d \mathbf{f}_d s_d + \mathbf{wH}_c \mathbf{f}_d s_c + \sum_{j=1}^L \mathbf{wH}_j \mathbf{f}_j s_j + {\bf wn},
    \label{eq:virtual_comm_rx}\\
    \text{SINR} &= \frac{|\mathbf{wH}_d \mathbf{f}_d|^2}{|\mathbf{wH}_c \mathbf{f}_d|^2 + \sum_{j=1}^L |\mathbf{wH}_j \mathbf{f}_j|^2 + \sigma^2 \|{\bf w}\|^2}.
    \label{eq:virtual_sinr}
\end{align}
By comparing these two formulations, it is observed that the expressions for the virtual SINR \eqref{eq:virtual_sinr} and the SCNR \eqref{eq:scnr} are identical. Since our objective is to guarantee a minimum SCNR, it is possible to address SCNR metric by transforming the radar link into a virtual communication link and defining the corresponding virtual SINR. This interpretation enables the application of the WMMSE algorithm to ensure that the minimum SCNR requirement is satisfied. 

The remaining task is to verify the convexity of the MSE for the virtual communication link with respect to ${\bf f}_d$. The MSE of the link \eqref{eq:virtual_comm_rx} is expressed as
\begin{align}
    \epsilon = \left|1 - \mathbf{w} \mathbf{h}_d \mathbf{f}_d \right|^2 + \left|\mathbf{w} \mathbf{h}_c \mathbf{f}_d \right|^2 + C, 
\end{align}
where $C = \sum_{j=1}^L \left|\mathbf{w} \mathbf{h}_j \mathbf{f}_j \right|^2 + \sigma^2 \|\mathbf{w}\|^2$ and it is straightforward to show that this expression remains convex with respect to ${\bf f}_d$, since the additional term $\left|\mathbf{w} \mathbf{h}_c \mathbf{f}_d \right|^2$ is also convex in ${\bf f}_d$.

Through this reinterpretation, it is demonstrated that communication-centric algorithms can be effectively extended to ISAC systems, thereby enlarging the applicability of the WMMSE algorithm to beamforming design under sensing performance constraints.


The WMMSE-based algorithm in \cite{shi2011iteratively} is adopted, with the aforementioned reinterpretation, to solve the sum rate maximization problem. A single-cell ISTN is interpreted as a two-cell multiuser MIMO system, where one cell consists of a LEO satellite and a SUT, and the algorithm is applied to this setup\footnotemark. 
\footnotetext{If $N_{\rm cell}\ge2$ terrestrial cells lie within each spot-beam footprint, the framework can be extended by modeling the ISTN downlink as a MIMO interfering broadcast channel with $N_{\rm cell}+1$ transmit nodes (one node is LEO satellite and $N_{\rm cell}$ nodes are TBSs) and by reusing the same algorithm.}
The WMMSE algorithm \cite{shi2011iteratively} consists of three steps, MMSE filtering, weight update, and transmitter update. To accommodate the QoS constraints on the SUT rate and SCNR, the third step is modified to solve an additional subproblem that ensures the required QoS constraints \cite{ma2016qos}.

The MSEs of the $k$th TUT, SUT, and the $i$th target are defined as follows.
\begin{align}
{\tilde \epsilon}_{k,m} &= \bigl|1-{\tilde {w}}_{k,m} {\bf h}_{m,k} {\bf f}_{m,k}\bigr|^2
+ \sum_{j\neq k}\bigl|{\tilde {w}}_{k,m} {\bf h}_{m,k} {\bf f}_{m,j}\bigr|^2 \nonumber\\
&+ \left|{\tilde {w}}_{k,m}\right|^2 g_{m,k}^{\rm TUT} p_m^{(\rm S)} + \sigma_{\rm n,T}^2 |{\tilde {w}}_{k,m}|^2,\\
{\bar \epsilon}_{m} &= \bigl|1-{\bar {w}}_{m}\sqrt{g_{m}^{\rm SUT} p_m^{(\rm S)}} \bigr|^2 + \sum_{j=1}^{K+N_{\rm tar}} \left|{{\bar {w}}_{m}\bf h}_{m}^{(\rm S)}{\bf f}_{m,j}\right|^2 \nonumber\\
&+ \sigma_{\rm n,S}^2 |{\bar {w}}_{m}|^2, \\
\epsilon_{i,m} &= \bigl|1-{\bf w}_{i,m} {\bf G}_{i,A(i,m),m}^{\rm tar} {\bf f}_{m,K+i}\bigr|^2 + I_{m,i}^{(\rm 1)} + I_{m,i}^{(\rm 2)} \nonumber\\
&+ I_{m,i}^{(\rm 3)} + I_{m,i}^{(\rm 4)}
+ \sigma_{\rm n,R}^2\|{\bf w}_{i,m}\|^2.
\end{align}

As in \cite{shi2011iteratively}, the joint beamforming and satellite power control algorithm consists of the following three steps:

1) \textbf{Receiver update (MMSE filter):}
\begin{align}
{\tilde {w}}_{k,m} &=
{\bf R}_{{\rm comm},m,k}^{-1} {\bf h}_{m,k} {\bf f}_{m,k}, \forall k \in [K],\\
{\bar {w}}_{m} &= \left({\bf h}_{m}^{(\rm S)} {\bf F}_m {\bf F}_m^{\rm H} ({\bf h}_{m}^{(\rm S)})^{\rm H} + \sigma_{\rm n,S}^2 \right)^{-1}\sqrt{g_{m}^{\rm SUT} p_m^{(\rm S)}},\\
{\bf w}_{i,m} &= {\bf R}_{{\rm sens},m,i}^{-1}  {\bf G}_{i,A(i,m),m}^{\rm tar} {\bf f}_{m,K+i}, \forall i \in [N_{\rm tar}], 
\end{align}
where $[K]$ denotes $[1,\cdots,K]$. ${\tilde {w}}_{k,m}$ and ${\bar w}_{m}$ are scalar receive equalizers for the $k$th TUT and the SUT inside the $m$th cell, respectively. ${\bf R}_{{\rm comm},m,k}$ and ${\bf R}_{{\rm sens},m,i}$ are defined as
\begin{align}
{\bf R}_{{\rm comm},m,k} &= {\bf h}_{m,k} {\bf F}_m {\bf F}_m^{\rm H} {\bf h}_{m,k}^{\rm H}
+ g_{m,k}^{\rm TUT} p_m^{(\rm S)}
+\sigma_{\rm n,T}^2, \\
{\bf R}_{{\rm sens},m,i} &= {\bf \bar G}_{m,A(i,m)} {\bf F}_m {\bf F}_m^{\rm H} {\bf \bar G}_{m,A(i,m)}^{\rm H} \nonumber\\
&+ {\bf g}_{m,A(i,m)}^{\rm rad} ({\bf g}_{m,A(i,m)}^{\rm rad})^{\rm H} p_m^{(\rm S)}
+\sigma_{\rm n,R}^2 {\bf I}_{N_{\rm RX}}.
\end{align}

2) \textbf{Weight update:} The WMMSE weight of the $k$th TUTs, SUTs, and the $i$th target are denoted by ${\tilde \mu}_{k,m}$, ${\bar \mu}_{m}$, and ${\mu}_{i,m}$. 
\begin{align}
{\tilde \mu}_{k,m} &= (1- {\tilde {w}}_{k,m} {\bf h}_{m,k} {\bf f}_{m,k})^{-1}, \forall k \in [K], \\
{\bar \mu}_{m} &= \left(1 - {\bar {w}}_{m}\sqrt{g_{m}^{\rm SUT} p_m^{(\rm S)}}\right)^{-1}, \\
{\tilde \mu}_{i,m} &= (1- {\bf w}_{i,m} {\bf G}_{i,A(i,m),m}^{\rm tar} {\bf f}_{m,K+i})^{-1}, \forall i \in [N_{\rm tar}].
\end{align}

3) \textbf{Transmitter update:}
Given fixed receiver beamforming vectors and weights, the transmit beamformer ${\bf F}_m$'s and satellite power $p_m^{(\rm S)}$'s are updated by solving
\begin{equation}\label{step3}
\begin{aligned}
&\mathop {\rm minimize}\limits_{{\bf F}_m, p_m^{(\rm S)}} \,\,\,\, \sum_{k=1}^{K} {\tilde \mu}_{k,m} {\tilde \epsilon}_{k,m} + {\bar \mu}_{m} {\bar \epsilon}_{m}\\
&\mathop {\rm subject\; to}\,\,\,\, {\bar \epsilon}_{m} \le 2^{-R_{\rm min}^{(\rm S)}}, \\
&\qquad\qquad\quad \epsilon_{i,m} \le \left({{\rm SCNR}_{\rm min}} +1\right)^{-1}, \forall i, \\
&\qquad\qquad\quad {\rm tr}\left({\bf F}_m^{\rm H} {\bf F}_m \right) \le P_{{\rm BS},m},
\end{aligned}
\end{equation}
where the minimum SUT rate and SCNR constraints are reformulated as maximum MSE constraints. The third step results in a convex optimization problem, which is efficiently solvable using CVX \cite{grant2014cvx}. The overall algorithm is solved by iterating these three steps until convergence.

The aforementioned algorithm can include both satellite rate and sensing constraints into well-known WMMSE algorithm, enlarging applicability of WMMSE algorithm with a simple extension.

\subsection{Merging Stage}
Next, in the merging stage, the power allocation for each spot beam should be centrally adjusted and enforced to satisfy the sum-power constraint of spot beams \eqref{constraint4}. The pre-optimized power allocation for the $m$th spot beam from (P1) is denoted as ${\bar p}_m$, which serves as a guideline for the $m$th terrestrial cell. Since the $m$th TBS will design its terrestrial beamforming based on the assumption that the $m$th spot beam will transmit with ${\bar p}_m$, the actual transmit power $p_m^{(\rm S)}$ should not exceed ${\bar p}_m$ to prevent unexpected excessive interference from the satellite.

The achievable sum rate maximization for $M$ spot beams with two constraints is considered as follows.
\begin{equation}\label{P2}
\begin{aligned}
{\rm P2} :\quad &\mathop {\rm maximize}\limits_{\{p_m^{(\rm S)}\}} \,\,\,\, \sum_{m=1}^M R_m^{(\rm S)} \\
&\mathop {\rm subject\; to}\,\,\,\, \sum_{m=1}^M p_m^{(\rm S)} \le P_{\rm LEO}, \\
&\qquad\qquad\quad p_m^{(\rm S)} \le {\bar p}_m, \quad \forall m. 
\end{aligned}
\end{equation}
The first constraint is identical to \eqref{constraint4}, which must be satisfied but has been difficult to enforce due to the distributed optimization approach. The second is the maximum per-beam power constraint. Problem \eqref{P2} is convex and can be solved with CVX \cite{grant2014cvx}.

\subsection{Refinement Stage}
At $\tau = \tau_0$, the TBSs estimate terrestrial CSI for the current time to design beamforming. In this process, the TBSs have prior knowledge that the satellite power will not exceed ${\bar p}_m$, as well as the resulting maximum interference. Therefore, the objective of this stage is to re-optimize the terrestrial beamforming matrices with the current terrestrial CSI under the worst-case assumption that $p_m^{(\rm S)} = {\bar p}_m, \forall m$.
\begin{equation}\label{P3}
\begin{aligned}
{\rm P3} :\quad &\mathop {\rm maximize}\limits_{{\bf F}_m} \,\,\,\, \sum_{k=1}^K R_{m,k}^{(\rm T)} + R_m^{(\rm S)} \\
&\mathop {\rm subject\; to}\,\,\,\, R_m^{(\rm S)} \ge R_{\rm min}^{(\rm S)}, \\
&\qquad\qquad\quad {\rm SCNR}_{i,m} \ge {\rm SCNR}_{\rm min}, \quad \forall i, \\
&\qquad\qquad\quad {\rm tr}\left({\bf F}_m^{\rm H} {\bf F}_m \right) \le P_{{\rm BS},m}.
\end{aligned}
\end{equation}
Note that the satellite power is not included as an optimization variable, since it has already been optimized via Algorithm 1 and Algorithm 2 and is given as a fixed constant. Problem \eqref{P3} can be solved with the same algorithm in Sec.~\ref{Sec:singlecell}, while treating the satellite power as a fixed parameter.

\subsection{Overall Algorithm}
For clarity, the overall procedure for executing the proposed algorithms is summarized below. For better understanding, we denote ${\bf H}_m[\tau]$, ${\bf g}_m [\tau]$, and $\tilde {\bf g}_m [\tau]$ as the terrestrial CSI, satellite CSI, and predicted satellite CSI for $m$th cell at time slot $\tau$, respectively.
\begin{figure}[h]
\centering
\subfloat[Algorithm workflow at $\tau = \tau_0 - \Delta\tau$.]{
  \includegraphics[trim={2.8cm 1.8cm 12.1cm 2.75cm},clip,width=0.94\columnwidth]{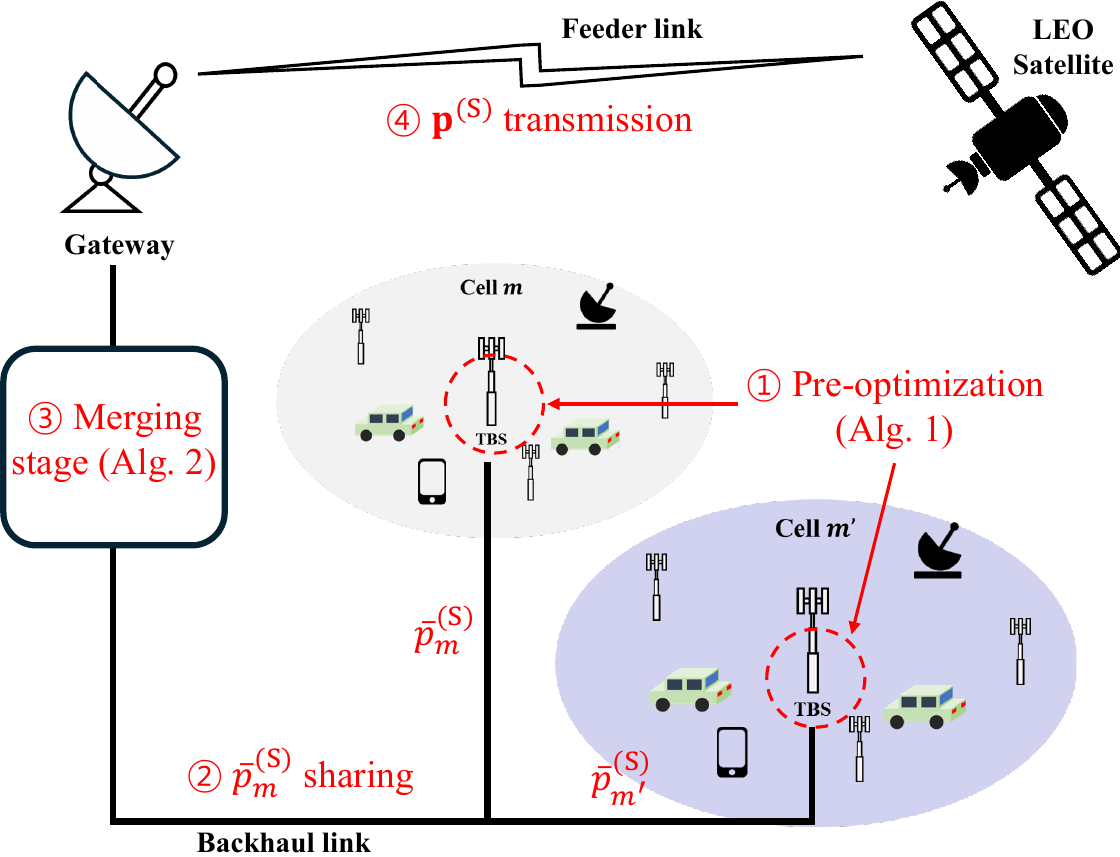}%
  \label{FIG:overall_1}
}

\subfloat[Algorithm workflow at $\tau = \tau_0$.]{
  \includegraphics[trim={5.4cm 1.8cm 12.1cm 2.75cm},clip,width=0.94\columnwidth]{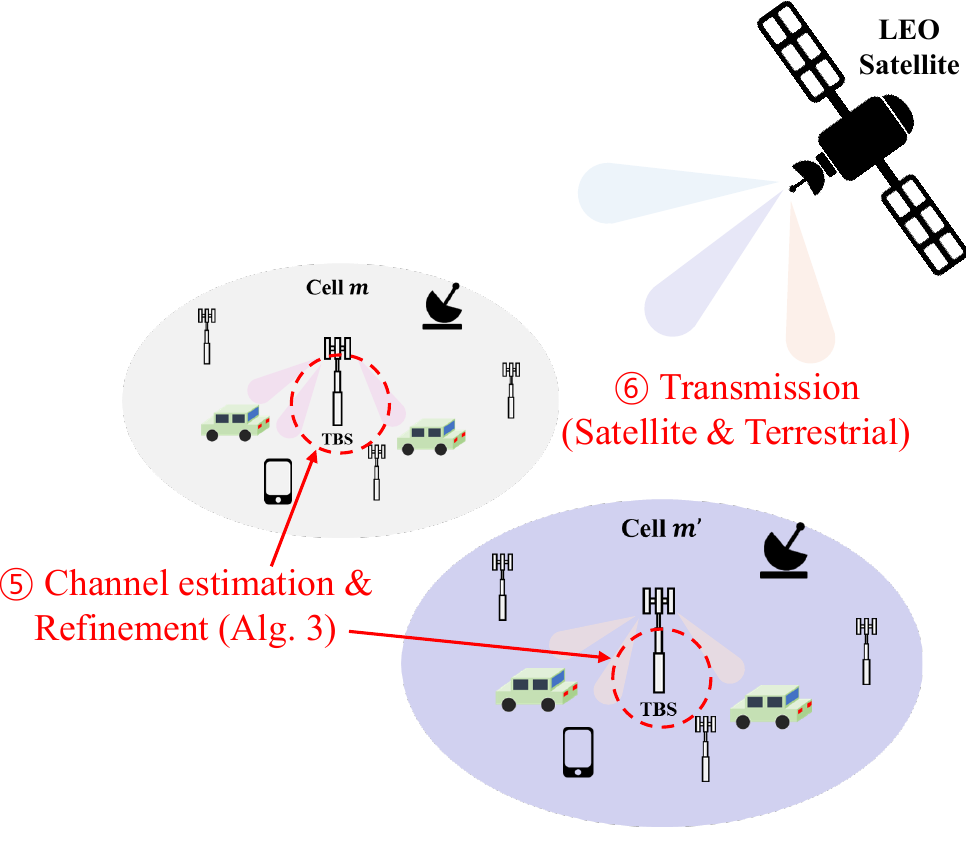}%
  \label{FIG:overall_2}
}
\caption{An illustration that shows overall algorithm workflow.}
\label{FIG:overall}
\end{figure}

Fig.~\ref{FIG:overall} illustrates the step-by-step operation of the proposed algorithm. 
In Fig.~\ref{FIG:overall}\subref{FIG:overall_1}, the operation at $\tau=\tau_0-\Delta\tau$ (pre-transmission preparation) is depicted. 
Based on $\tilde {\bf g}_m [\tau_0]$ (predicted from ${\bf g}_m [\tau_0-\Delta\tau]$ and orbital prediction) and ${\bf H}_m[\tau_0 - \Delta\tau]$, each TBS executes the pre-optimization algorithm to obtain the optimal satellite power for the $m$th cell, i.e., ${\bar p}_m^{(\mathrm{S})}$. 
These ${\bar p}_m^{(\mathrm{S})}$'s are shared over the backhaul link, and the processing unit executes the merging algorithm to produce the vector $\mathbf{p}^{(\mathrm{S})}$. 
Subsequently, $\mathbf{p}^{(\mathrm{S})}$ is sent to the LEO satellite via the feeder link.

Fig.~\ref{FIG:overall}\subref{FIG:overall_2} illustrates the operation at $\tau=\tau_0$. 
Each TBS estimates current CSI ${\bf H}_m[\tau_0]$. 
Given the known satellite interference (from the predefined ${\bar p}_m^{(\mathrm{S})}$'s), each TBS computes its beamforming matrix via the refinement algorithm. 
Then, each TBS transmits the ISAC signal using the computed beamforming matrix, while the LEO satellite transmits data streams for the SUTs based on the predefined power vector $\mathbf{p}^{(\mathrm{S})}$.

Ideally, both terrestrial beamforming and satellite power allocation should be updated every coherence block by the pre-optimization, merging, and refinement stages. 
In practice, given that the satellite large-scale channel varies slowly compared to the terrestrial channel, only the terrestrial beamforming is updated at each terrestrial coherence time via the refinement stage with fixed ${\bar p}_m^{(\mathrm{S})}$'s, and the satellite power allocation is recomputed via pre-optimization and merging stages only upon significant large-scale changes. Under the proposed framework, each TBS only needs to share ${\bar p}_m^{(\mathrm{S})}$ rather than full CSI, thereby alleviating the cooperation overhead.

\section{Simulation Results}
In this section, the performance of the proposed techniques under ISTN settings is evaluated by simulations.

\subsection{Simulation Settings}

The LEO satellite is assumed to operate at an altitude of 500 km, with its elevation angle randomly selected within $60^{\circ}$. Each spot beam covers a footprint with a radius of 10 km. The carrier frequency is set to $f_c = 15 \,{\rm GHz}$ (Ku-band), and the system bandwidth is ${\rm BW} = 100 \,{\rm MHz}$. The receiver noise parameters are given by a noise figure of $N_F = 8 \,{\rm dB}$ and a noise spectral density of $N_0 = -174 \,{\rm dBm/Hz}$. 
Other satellite parameters are configured as $\sigma_{\xi} = 4 \,{\rm dB}$, satellite antenna gain $G_s = 38 \,{\rm dBi}$, receiver main-lobe gain $G_{\rm main} = 34.2 \,{\rm dBi}$, receiver side-lobe gain $G_{\rm side} = 21.2 \,{\rm dBi}$, and half-power beamwidth $\varphi_{\rm 3dB} = 1^{\circ}$~\cite{kim2024spectrum,choi2023joint}. The maximum total transmit power of the satellite is limited to $P_{\rm LEO} = 200 \,{\rm W}$. 

For the terrestrial network, four cells are considered, each located within the footprint of one spot beam. The radius of each cell is $200 \,{\rm m}$. The TBS and TUT heights are set to $10 \,{\rm m}$ and $1.5 \,{\rm m}$, respectively. Each TBS is placed at the center of its cell, while TUTs are uniformly distributed. The system parameters are specified as $N_{\rm TX} = 30$, $K = 5$, $N_{\rm RX} = 4$, and $N_{\rm tar} = 4$. Propagation characteristics such as path loss, shadowing, and LOS probability follow the QuaDRiGa simulator with the 3GPP TR 38.901 urban macrocell channel model \cite{burkhardt2014quadriga,3gpp.38.901}. 

Unless otherwise noted, sensing targets are uniformly distributed within a radius of $r_{\rm sens} = 50\,{\rm m}$ inside each cell. Radar receivers are located at $(\frac{r_{\rm sens}}{2},0)$, $(0,\frac{r_{\rm sens}}{2})$, $(-\frac{r_{\rm sens}}{2},0)$, and $(0,-\frac{r_{\rm sens}}{2})$. The number of clutters per target is set to $N_{\rm cl} = 3$.

\subsection{Performance Evaluation}

First, the achievable sum rate and SCNR of the proposed and baseline techniques are evaluated in a single terrestrial cell with a single-beam LEO satellite, i.e., $M=1$. The minimum SCNR constraint is set to ${\rm SCNR}_{\rm min} = -10\,{\rm dB}$. 

For performance comparison, several baselines are considered as follows. Unless otherwise specified, the satellite allocates transmit power based on the water-filling algorithm and each TBS employs zero-forcing (ZF) beamforming, allocates power to satisfy the SCNR constraint, and distributes the residual power to TUTs using water-filling. The target–receiver association is assumed to follow a nearest-neighbor method, where each target is connected to the receiver with the highest channel gain.

\textbf{(i) Interference-free} represents an ideal upper bound in which each system causes no interference to the other.
\textbf{(ii) Zero-forcing with exhaustive power search (ZF-EPS)} refers to the method that finds the optimal power allocation for the satellite and TBS via exhaustive search to maximize the achievable sum rate, assuming ZF beamforming is employed.
\textbf{(iii) Greedy} corresponds to the case where the satellite and terrestrial networks independently maximize their own performance metrics while ignoring the interference imposed on the other system. 
\textbf{(iv) Uniform} allocates transmit power for TUTs uniformly, disregarding inter-system interference.

\begin{figure}[h]
    \centerline{\includegraphics[trim={0.1cm 0.12cm 0.1cm 0.2cm},clip,width=0.94\linewidth]{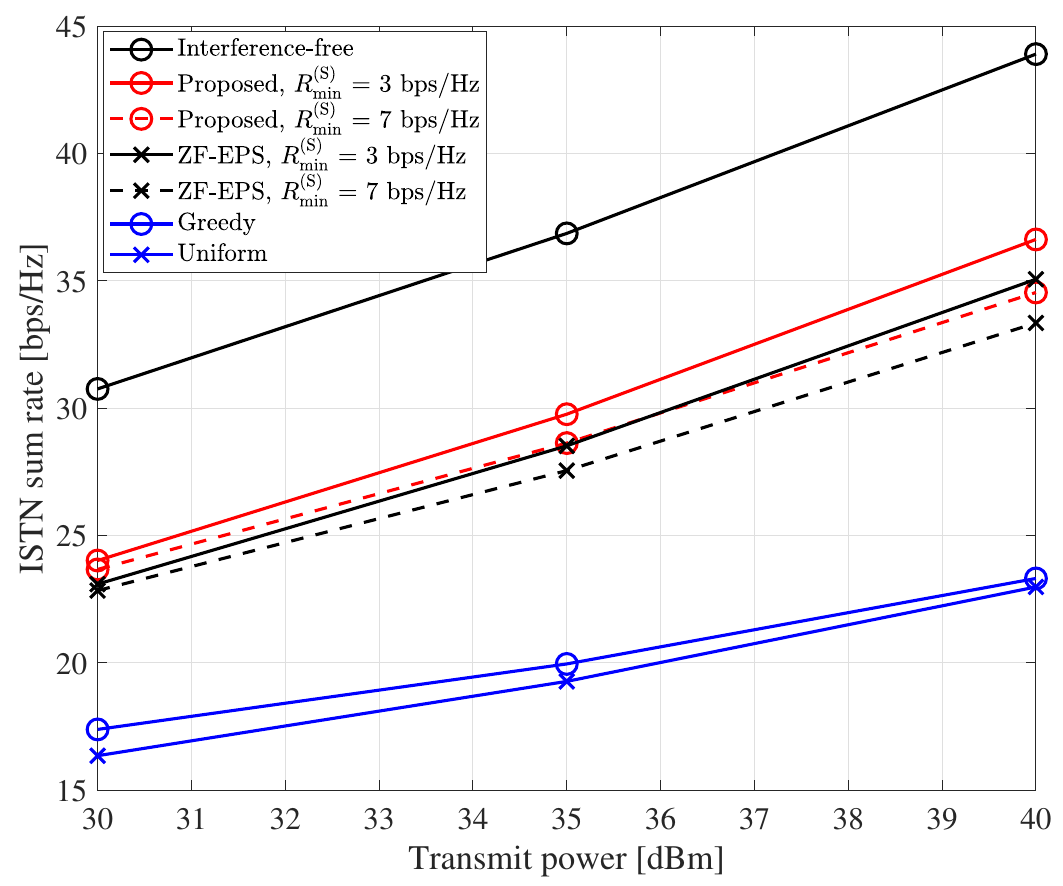}}
    \caption{ISTN sum rate versus $P_{{\rm BS},m}$: ${\rm SCNR}_{\rm min} = -10\,{\rm dB}$.}
    \label{FIG:SR_concat} 
\end{figure}

Fig. \ref{FIG:SR_concat} shows the ISTN sum rate, defined as the sum of the terrestrial and satellite sum rates, with respect to $P_{{\rm BS},m}$ for $R_{\rm min}^{(\rm S)} = 3\,{\rm bps/Hz}$ and $7\,{\rm bps/Hz}$.
The greedy method suffers from significant rate loss due to mutual interference, about $50\%$ relative to the ideal baseline although the same technique with Interference-free case is applied. This indicates that a sophisticated design is required for systems with shared spectrum. Furthermore, when $R_{\rm min}^{(\rm S)}$ increases from $3\,{\rm bps/Hz}$ to $7\,{\rm bps/Hz}$, the achievable rate of both the proposed and ZF-EPS further decreases. This is because the satellite employs higher transmit power to guarantee the stricter SUT rate, which in turn causes stronger interference to the terrestrial network. The proposed method outperforms the baselines by employing a better association and beamforming strategy that explicitly accounts for satellite interference. Although ZF-EPS finds the optimal power allocation for both satellite and TBS, the proposed method can achieve additional gain with a better design of beamforming and user association for suppressing inter-system interference.

\begin{figure}[h]
    \centerline{\includegraphics[trim={0.1cm 0cm 0.1cm 0.1cm},clip,width=0.94\linewidth]{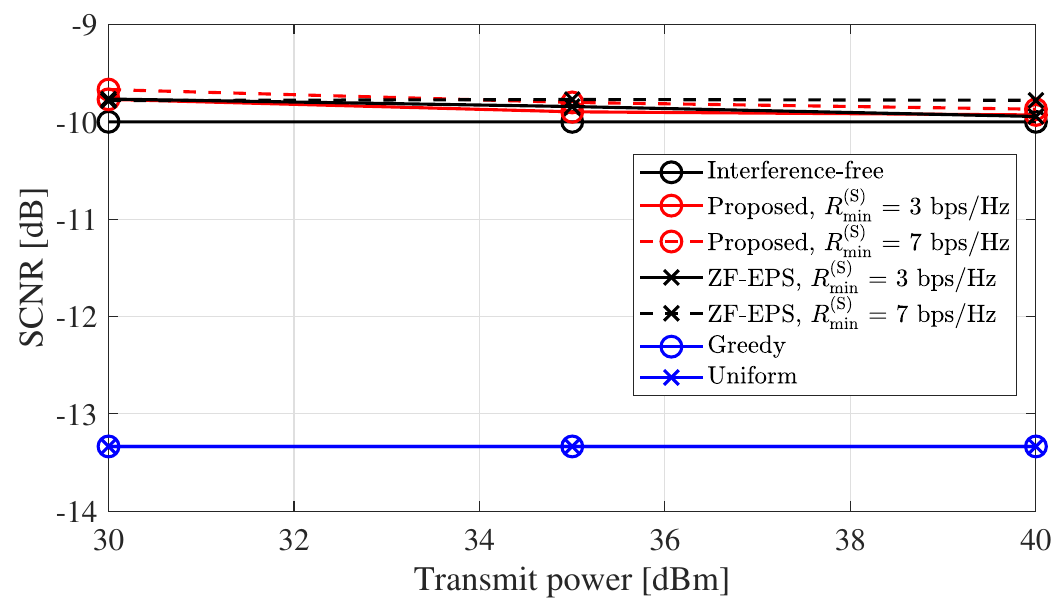}}
    \caption{SCNR versus $P_{{\rm BS},m}$: ${\rm SCNR}_{\rm min} = -10\,{\rm dB}$.}
    \label{FIG:SCNR_concat} 
\end{figure}

Fig. \ref{FIG:SCNR_concat} shows the SCNR of targets with the same settings as in Fig.~\ref{FIG:SR_concat}. The proposed method is observed to consistently guarantee a SCNR higher than $-10\,{\rm dB}$. In contrast, the Greedy and Uniform baselines allocate transmit power to meet the minimum SCNR constraint, however, additional SCNR degradation occurs since they ignore interference from the satellite.

\begin{figure}[h]
\centering
\subfloat[$R_{\rm min}^{(\rm S)} = 3\,{\rm bps/Hz}$.]{
  \includegraphics[trim={0.1cm 0.12cm 0.1cm 0.2cm},clip,width=0.472\columnwidth]{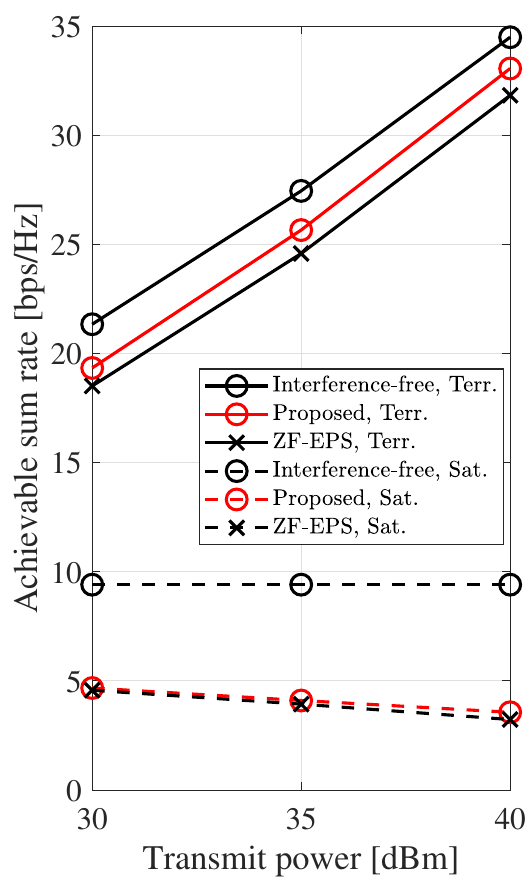}%
  \label{FIG:TS_rate3}
}
\hfil
\subfloat[$R_{\rm min}^{(\rm S)} = 7\,{\rm bps/Hz}$.]{
  \includegraphics[trim={0.1cm 0.12cm 0.1cm 0.2cm},clip,width=0.472\columnwidth]{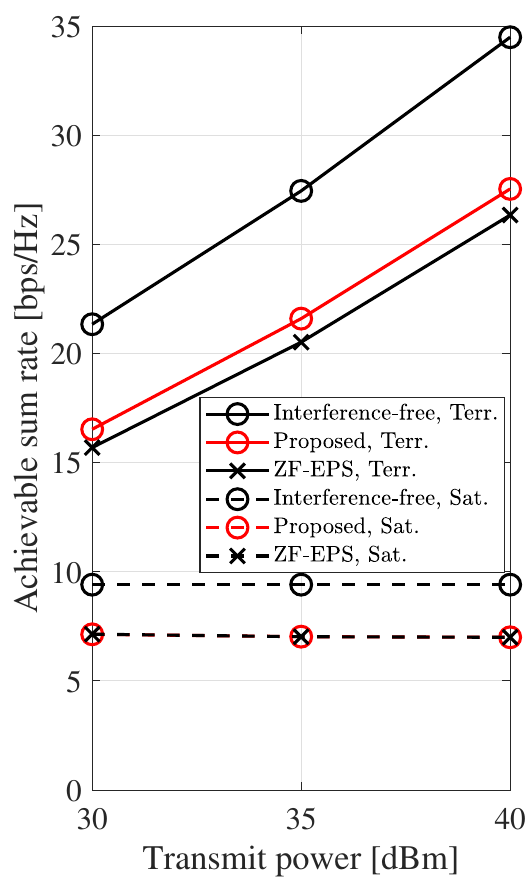}%
  \label{FIG:TS_rate7}
}
\caption{The terrestrial and satellite sum rate performance:
${\rm SCNR}_{\rm min} = -10\,{\rm dB}$.}
\label{FIG:TSrate}
\end{figure}

Fig.~\ref{FIG:TSrate} shows the achievable sum rates of the terrestrial and satellite systems separately. As expected, larger terrestrial rate loss occurs when the SUT rate requirement $R_{\rm min}^{(\rm S)}$ is higher. In Fig.~\ref{FIG:TSrate}\subref{FIG:TS_rate3}, compared to the interference-free case, most of the ISTN rate loss arises from the decreased satellite rate, while the terrestrial rate remains comparable to the ideal case. In contrast, in Fig.~\ref{FIG:TSrate}\subref{FIG:TS_rate7}, the majority of ISTN rate loss originates from the terrestrial rate rather than the satellite rate, indicating that interference from the satellite becomes the dominant factor. 
It is also observed that the optimized satellite rate in Fig.~\ref{FIG:TSrate}\subref{FIG:TS_rate3} is larger than $R_{\rm min}^{(\rm S)} = 3\,{\rm bps/Hz}$, whereas in Fig.~\ref{FIG:TSrate}\subref{FIG:TS_rate7}, it is almost the same as $R_{\rm min}^{(\rm S)} = 7\,{\rm bps/Hz}$. This implies that increasing the satellite rate beyond $R_{\rm min}^{(\rm S)} = 3\,{\rm bps/Hz}$ is beneficial in perspective of the overall ISTN sum rate. However, increasing the satellite rate beyond $R_{\rm min}^{(\rm S)} = 7\,{\rm bps/Hz}$ becomes detrimental to ISTN sum rate.

Next, we validate the effectiveness of the proposed association technique by comparing it with several baselines. In the Nearest Neighbor association, each target is assigned to the receiver that yields the largest channel gain. In the Greedy association, targets are sequentially associated with the receiver providing the largest channel gain, under the constraint that each receiver can serve at most one target. In the Random association, targets and receivers are paired in a random one-to-one mapping without considering channel conditions.

\begin{figure}[h]
\centering
\subfloat[$N_{\rm RX}=4$.]{
  \includegraphics[trim={0.1cm 0.12cm 0.1cm 0.2cm},clip,width=0.472\columnwidth]{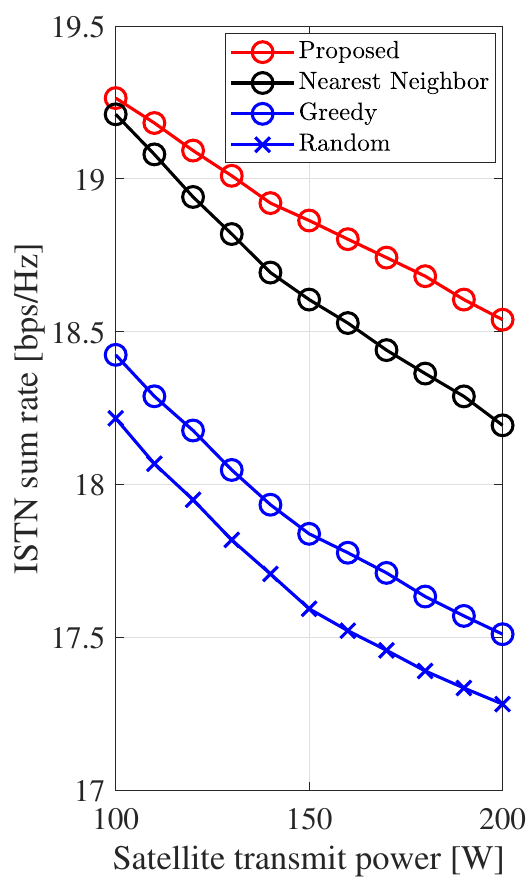}%
  \label{FIG:Asso_N4}
}
\hfil
\subfloat[$N_{\rm RX}=10$.]{
  \includegraphics[trim={0.1cm 0.12cm 0.1cm 0.2cm},clip,width=0.472\columnwidth]{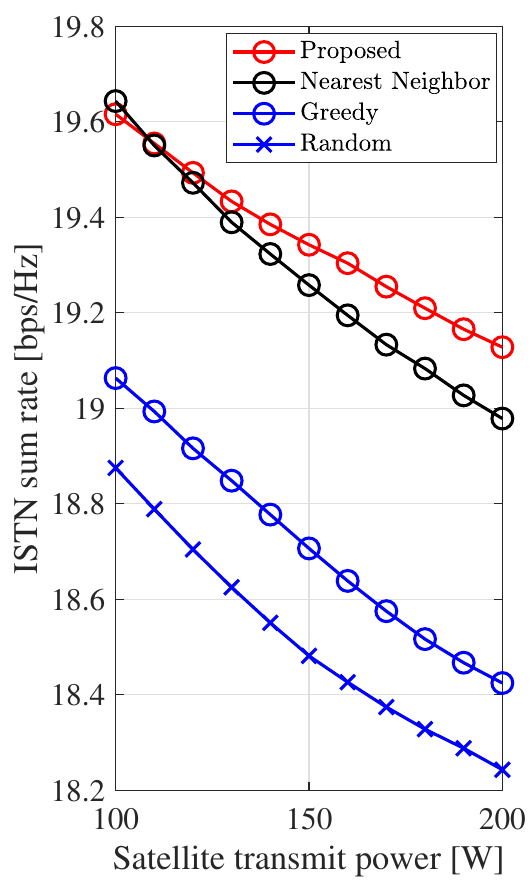}%
  \label{FIG:Asso_N10}
}
\caption{ISTN sum rate with different association methods: ${\rm SCNR}_{\rm min} = -10 \,{\rm dB}$.}
\label{FIG:Association}
\end{figure}

Fig.~\ref{FIG:Association} shows the achievable sum rate with different association techniques. As the satellite transmit power increases, the achievable rate decreases for all association methods due to stronger interference. Among the methods, the proposed association achieves the best performance by explicitly considering the satellite direction when matching targets to radar receivers. Although the Nearest Neighbor association connects each target to the receiver with the largest channel gain, it does not consider interference from other targets or satellites, resulting in inferior sum rate compared to the proposed method.

Comparing Fig.~\ref{FIG:Association}\subref{FIG:Asso_N4} and Fig.~\ref{FIG:Association}\subref{FIG:Asso_N10}, the performance gap between the proposed association and the baselines becomes larger with smaller $N_{\rm RX}$. This is because, with fewer transmit antennas, it becomes more difficult to suppress satellite interference using receive beamforming only. 
These results indicate that the proposed association is relatively advantageous when the number of receive antennas per radar is small, suggesting a potential benefit in multistatic ISAC systems, where the same total number of radar antennas are distributed across multiple radar units.

To further investigate the impact of satellite signals on sensing performance, we evaluate the sensing failure probability with respect to the sensing cell radius $r_{\rm sens}$, defined as the radius of the area in which sensing targets exist. The sensing failure probability denotes the probability that the SCNR constraint cannot be satisfied even if the full TBS transmit power is dedicated for sensing.

\begin{figure}[h]
    \centerline{\includegraphics[trim={0.1cm 0.12cm 0.1cm 0.2cm},clip,width=0.94\linewidth]{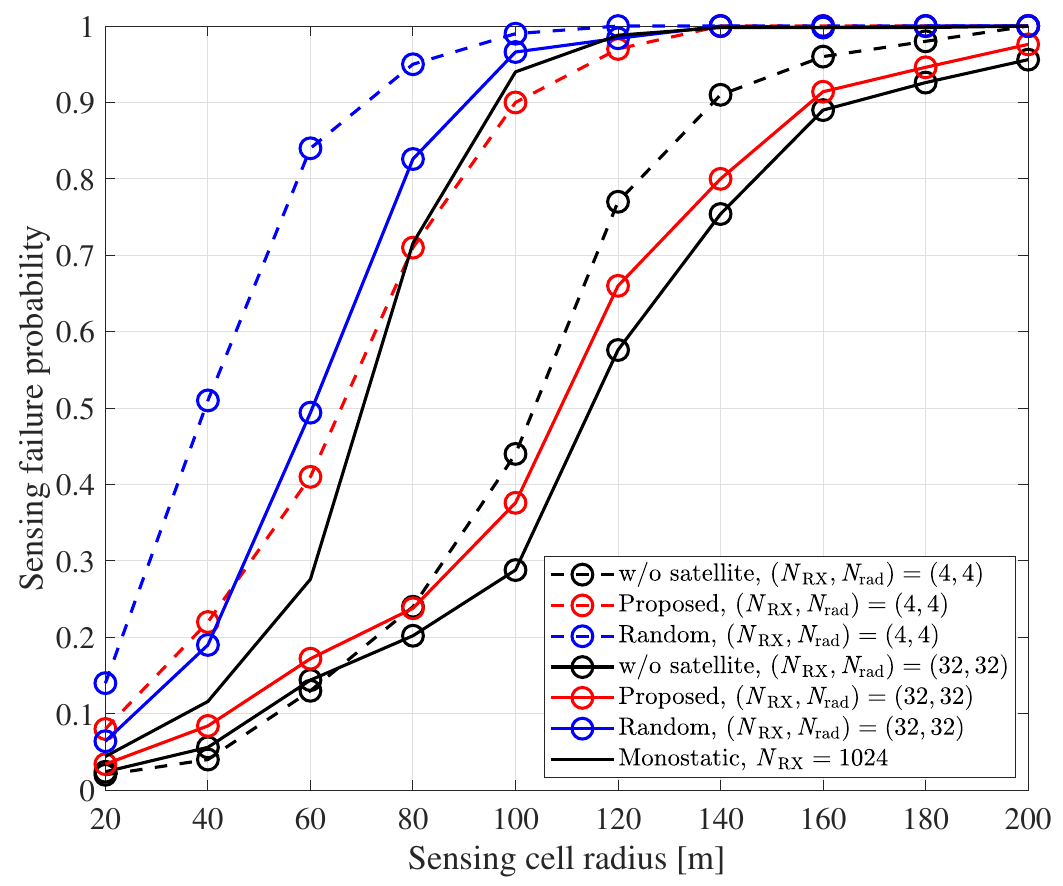}}
    \caption{The sensing failure probability with respect to cell radius.}
    \label{FIG:fail_prob_N} 
\end{figure}

In Fig.~\ref{FIG:fail_prob_N}, the sensing failure probability of different cases is presented. As the sensing cell radius increases, the average path-loss becomes larger, resulting in higher failure probability. When $(N_{\rm RX},N_{\rm rad})=(4,4)$, the proposed method achieves a sensing failure probability of $0.3$ at approximately $44\%$ smaller radius than the satellite-free case, corresponding to about a $65\%$ reduction in the sensing coverage area. Meanwhile, the Random association case suffers more severe degradation, requiring nearly $67\%$ smaller radius, about $89\%$ reduction in area to maintain the same failure probability. These results indicate that while the presence of satellites inevitably reduces the visible sensing range, the proposed method can significantly mitigate this degradation compared to naive association techniques. 

For $(N_{\rm RX},N_{\rm rad})=(32,32)$, the proposed method yields a sensing failure probability that closely approaches the ideal satellite-free case, while the Random association still shows large failure probability even with a large number of receive antennas. This demonstrates that even with abundant antenna resources, effective sensing cannot be achieved under satellite interference without an appropriate association method. Furthermore, it is observed that the monostatic ISAC configuration cannot outperform the proposed multistatic setup, despite using the same total number of receive antennas, thereby highlighting the superiority of multistatic ISAC systems.

\begin{figure}[h]
    \centerline{\includegraphics[trim={0.1cm 0.12cm 0.1cm 0.2cm},clip,width=0.94\linewidth]{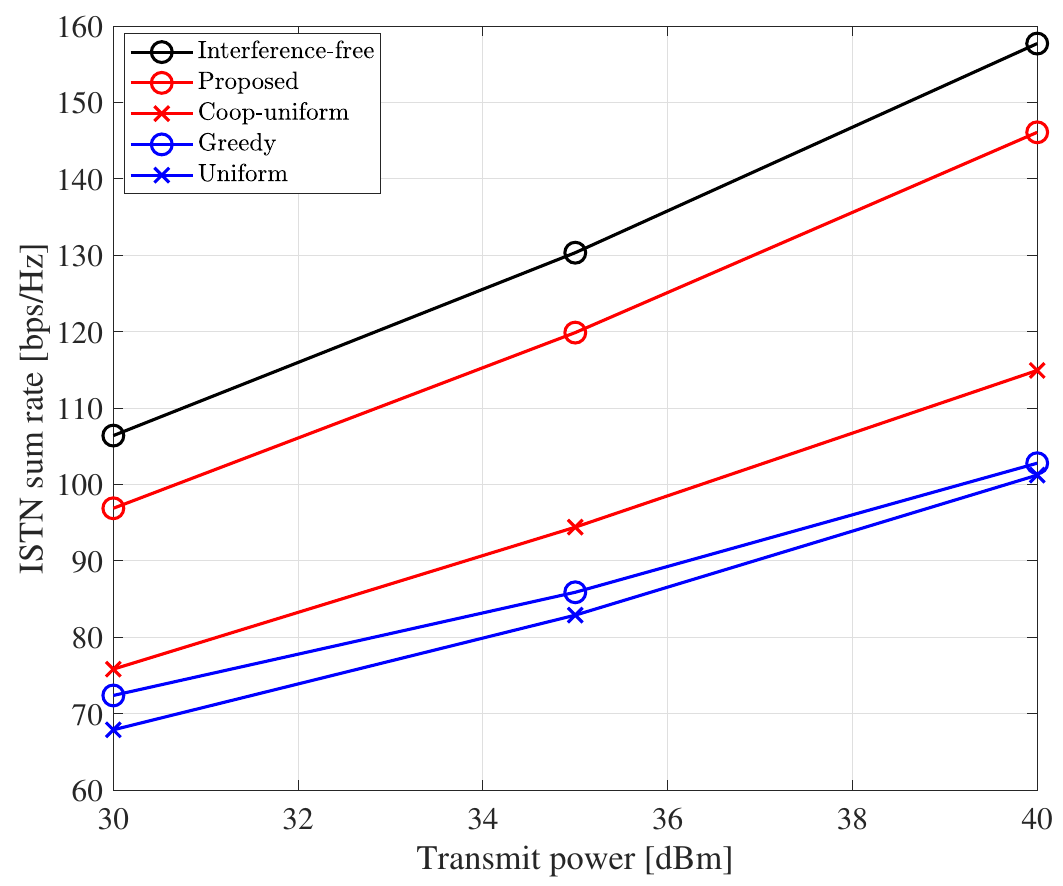}}
    \caption{ISTN sum rate of multicell systems: $M=4$, ${\rm SCNR}_{\rm min} = -10 \,{\rm dB}$, and ${R}_{\rm min}^{(\rm S)} = 5 \,{\rm bps/Hz}$.}
    \label{FIG:sumrate_multicell} 
\end{figure}

Fig.~\ref{FIG:sumrate_multicell} shows the ISTN sum rate when $M=4$. Additional baseline Coop-uniform denotes the case that TBSs cooperatively design beamforming to maximize the ISTN sum rate, while the satellite uses uniform power allocation. Compared with the single-cell case ($M=1$) in Fig.~\ref{FIG:SR_concat}, the performance gap between the ideal baseline and the proposed method becomes smaller. The reason for this result can be explained as follows. Excessive power usage by the satellite degrades the terrestrial network rate, and when $M=1$, the satellite cannot fully exploit its entire power budget in order to avoid severe interference. In contrast, when $M\ge2$, the optimal satellite power of each spot beam varies depending on the terrestrial CSI. Consequently, the satellite can fully exploit its available power and allocate it more efficiently to spot beams that offer higher gains while causing less interference to terrestrial networks, thereby enhancing the overall sum rate. Although terrestrial beamforming is designed by accounting for inter-system interference, the Coop-uniform baseline cannot approach the performance of the ideal baseline, as the satellite employs a fixed, non-adaptive power allocation strategy.

\begin{figure}[h]
    \centerline{\includegraphics[trim={0.1cm 0.12cm 0.1cm 0.2cm},clip,width=0.94\linewidth]{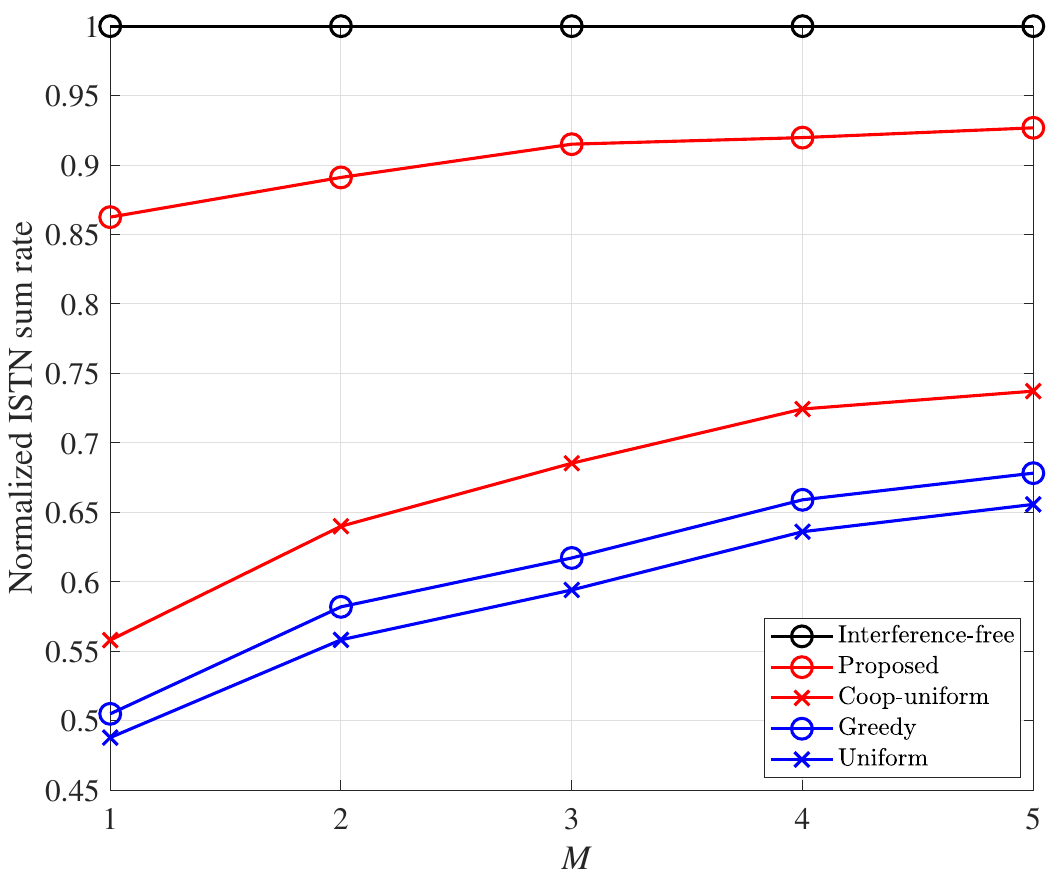}}
    \caption{Normalized ISTN sum rate with respect to $M$: ${\rm SCNR}_{\rm min} = -10 \,{\rm dB}$, $P_{{\rm BS},m} = 35 \,{\rm dBm}$, and ${R}_{\rm min}^{(\rm S)} = 5 \,{\rm bps/Hz}$.}
    \label{FIG:sumrate_multicell_M} 
\end{figure}
Fig.~\ref{FIG:sumrate_multicell_M} shows the ISTN sum rate normalized by that of the interference-free baseline, with respect to $M$. The proposed method achieves a sum rate greater than $90 \%$ of the ideal baseline when $M=5$. As discussed above, increasing the number of spot beams $M$ increases the degree-of-freedom for satellite power allocation, which remarkably reduces the ISTN sum rate gap from the perspective of the integrated network.

Overall, the simulation results indicate that multistatic ISAC–based integrated coexistence between satellite and terrestrial networks is feasible with only a modest performance penalty compared with the idealized interference-free case, provided that a sophisticated cooperation method, a large number of spot beams, and a large number of radar receivers and antennas are available.

\section{Conclusions}

This paper has presented the integrated coexistence framework between a multi-beam LEO satellite network and a terrestrial network that performs multistatic ISAC. To address the channel aging issue and enable low-complexity joint optimization across distributed cells, the distributed optimization framework for ISTNs has been proposed. For each cell, the joint beamforming and satellite power allocation method has been designed to maximize the achievable ISTN sum rate while ensuring minimum SCNR and satellite rate constraints, based on applicability of WMMSE algorithm for ISAC systems. In addition, the target–receiver association algorithm has been developed to mitigate directional interference from the LEO satellite. Simulation results confirm that the proposed approach significantly reduces the terrestrial performance degradation caused by satellite interference, thereby demonstrating the feasibility of terrestrial–satellite coexistence in the same frequency band.

Furthermore, two main conditions have been identified that allow ISTN performance to approach the ideal interference-free scenario. First, as the number of spot beams supported by a LEO satellite increases, the rate gap with respect to the ideal case decreases with a better opportunity to utilize satellite power in more advantageous manner. Second, as the number of radar receivers and receiver antennas increases, the sensing performance also approaches the interference-free benchmark. These findings provide useful design guidelines for future ISTNs, and a more in-depth investigation into the conditions that enable ISTNs to achieve near-optimal performance, along with a detailed analysis of the performance gap, will be pursued in future research. Future research may also consider the design of cooperative strategies that enable LEO satellites to actively assist the ISAC functionality of ISTNs, moving beyond their role as competitive sources, as well as more advanced problems arising in multi-satellite constellation scenarios.

\bibliographystyle{IEEEtran} 
\bibliography{IEEEabrv,Jee_2025_ISTN}      

\vfill

\end{document}